# Reactive single-step hot-pressing and magnetocaloric performance of polycrystalline Fe₂Al$_{1.15-x}$B₂Ge$_x$Ga$_x$ (x=0, 0.05) MAB phases

Benedikt Beckmann[1,*,+], Tarek A. El-Melegy[2,+], David Koch[1], Ulf Wiedwald[3], Michael Farle[3], Fernando Maccari[1], Joshua Snyder[2], Konstantin P. Skokov[1], Michel W. Barsoum[2], and Oliver Gutfleisch[1]

[1]Institute of Materials Science, Technical University of Darmstadt, Darmstadt 64287, Germany
[2]Department of Materials Science & Engineering, Drexel University, Philadelphia, PA, USA
[3]Faculty of Physics and Center for Nanointegration Duisburg-Essen (CENIDE), University of Duisburg-Essen, 47048 Duisburg, Germany

[+]Authors contributed equally
[*]Corresponding author
benedikt.beckmann@tu-darmstadt.de (B. Beckmann)
tae39@drexel.edu (T.A. El-Melegy)

## Abstract

Reactive single-step hot-pressing at 1473 K and 35 MPa for 4 h produces dense, bulk, near single-phase, low-cost and low-criticality Fe₂Al$_{1.15}$B₂ and Fe₂Al$_{1.1}$B₂Ge$_{0.05}$Ga$_{0.05}$ MAB samples, showing a second-order magnetic phase transition with favorable magnetocaloric properties around room temperature. The magnetic as well as magnetocaloric properties can be tailored upon Ge and Ga doping, leading to an increase of Curie temperature $T_C$ and spontaneous magnetization $m_S$. The maximum isothermal entropy change $\Delta s_{T,max}$ of hot-pressed Fe₂Al$_{1.15}$B₂ in magnetic field changes of 2 and 5 T amounts to 2.5 and 5 J(kgK)$^{-1}$ at 287.5 K and increases by Ge and Ga addition to 3.1 and 6.2 J(kgK)$^{-1}$ at 306.5 K, respectively. The directly measured maximum adiabatic temperature change $\Delta T_{ad,max}$ is improved by the composition modification from 0.9 to 1.1 K in magnetic field changes of 1.93 T. Overall, we demonstrate that hot-pressing provides a much faster, more scalable and processing cost reducing alternative compared to conventional synthesis routes to produce heat exchangers for magnetic cooling devices. Therefore, our criticality assessment shows that hot-pressed Fe-based MAB phases provide a promising compromise of material and processing cost, criticality and magnetocaloric performance, demonstrating the potential for low-cost and low-criticality magnetocaloric applications around room temperature.

# 1. Introduction

Energy-efficient and environmentally friendly cooling technology is imperative to face the rising energy demand for cooling [1] as well as the interconnected challenges of climate change, population growth and world-wide rising standard of living [2]. Today, most cooling devices are still based on the vapor-compression cycle, which lacks not only in thermodynamic efficiency [3] but is also environmentally damaging [4]. Magnetic refrigeration is considered to be a promising, environmentally friendly [5] alternative to conventional cooling technology, which shows a higher thermodynamic efficiency [5,6]. This solid-state cooling technology is based on the magnetocaloric effect, which is defined as the isothermal entropy change $\Delta s_T$ and adiabatic temperature change $\Delta T_{ad}$ upon application of an external magnetic field as stimulus [7].

There is a large variety of magnetocaloric materials [8,9], showing first- and/or second-order phase transitions. Materials with first-order phase transitions, such as $Gd_5(SiGe)_4$ [10], Fe-Rh [11,12], Ni-Mn-based Heusler alloys [13–15], $La(Fe,Si)_{13}$-type [16–18] and $Fe_2P$-type [19,20] compounds, show large magnetocaloric effects due to field-induced magnetostructural transitions. However, they suffer from hysteresis losses [21–24], degradation caused by mechanical failure due to large volume changes [20,25] and reduced magnetocaloric effects during cyclic magnetic field application [26,27]. Materials with second-order phase transitions, specifically the magnetic transition from ferro- to paramagnetic state at the Curie temperature $T_C$, are for instance Gd [28], Gd-Y [8], high entropy transition metal $NiFeCoCrPd_{0.5}$ alloys [29], Ni-Mn-based Heusler alloys [30,31] and $Fe_2AlB_2$ [32,33]. It should be noted that Ni-Mn-based Heusler alloys can show both magnetostructural martensitic first-order phase transitions [13–15] and magnetic second-order phase transitions at $T_C$ of the ferromagnetic phase [30,31]. Due to the absence of large volume changes and hysteresis, second-order phase transition materials show highly stable and reversible magnetocaloric effects. Based on these benefits, high-purity Gd is currently the benchmark material for magnetocalric devices operating at room temperature [34,35]. However, the high-cost and high-criticality of Gd [8,36–40] make its widespread commercial usage unattainable. Therefore, in this study, we have selected the low-cost, low-criticality and thereby more scalable $Fe_2AlB_2$-type MAB phases, showing promising magnetocaloric properties for room temperature application [32,33].

MAB phases are ternary transition metal borides that crystallize in the orthorhombic crystal system, where M is a transition metal, A is aluminum and B is boron. They have a variety of chemical formulas, namely: MAlB, $M_2AlB_2$, $M_3AlB_4$, $M_4AlB_6$ and $M_4AlB_4$ [41]. The crystal structure of $Fe_2AlB_2$ (space group *Cmmm*) consists of Fe-B blocks interleaved by pure Al layers and was first reported by W. Jeitscko [42] and Y. Kuzma *et al.* [43]. The $Fe_2AlB_2$ MAB phase formation is based on a peritectic reaction between an Al-rich liquid and FeB at around 1553 K [44]. To suppress the formation of FeB impurity phase, Al excess is added during synthesis, however, it in turn promotes the formation of $Al_{14}Fe_4$ impurities [45]. Both impurity phases interfere with and contribute to the overall magnetic and magnetocaloric behavior of $Fe_2AlB_2$-type compounds [44].

Previous studies on $Fe_2AlB_2$ show that the magnetism is of itinerant character [46–48], the saturation magnetization at $T \leq 50$ K measures $0.7 - 1.60$ $\mu_B$ / Fe atom [44–52] and the Curie temperature $T_C$ lies between $272 - 320$ K [44–55], indicating a second-order magnetic transition [45,50,51,53,56] around room temperature from the low-temperature ferromagnetic to the high-temperature paramagnetic state. Moreover, bulk $Fe_2AlB_2$ shows soft magnetic behavior with negligible remanence and coercivity [41,49,57,58]. However, due to its highly anisotropic crystal structure, $Fe_2AlB_2$ single crystals display highly anisotropic magnetic properties. Lamichhane *et al.* ranked the magnetization axes from the easiest to the hardest at 2 K to be [100], [010] and [001], respectively [46], with previous studies supporting this observation [48,50].

Based on its second-order magnetic transition, Fe$_2$AlB$_2$ shows a reversible magnetocaloric effect without cracking and hysteresis losses near room temperature [45,49,50,52]. The isothermal entropy change $\Delta s_T$ and indirectly determined adiabatic temperature change $\Delta T_{ad}$ are in the range of 1.3 – 4.4 J(kgK)$^{-1}$ [45,46,48–50,52,55] and 1.0 – 1.8 K [48,49] at a magnetic field change of 2 T, respectively. In addition, it has been shown that the compound shows favorable secondary functionalities, such as an anisotropic thermal conductivity determined in textured polycrystals at room temperature with minimum and maximum values of 8.5 and 10.6 W(Km)$^{-1}$ [59], respectively, providing fast thermal response and good heat exchange.

Recently, Barua *et al.* have demonstrated the tunability of the magnetocaloric effect in Fe$_2$AlB$_2$ through Ga and Ge doping via suction casting and subsequent annealing at 1373 K for 72 h [48]. The doping leads to an increase in $\Delta s_T$ and indirectly determined $\Delta T_{ad}$ to 6.5 J(kgK)$^{-1}$ and 2.2 K in magnetic field changes of 2 T, respectively. These enhancements are attributed to chemical bonding and electronic structure modifications by antisite occupancy of Fe on the Al-layer within the lattice, influencing the magnetic and magnetocaloric properties.

In this study, we continue the investigation of Ge and Ga doping in Fe$_2$AlB$_2$ by synthesizing Fe$_2$Al$_{1.15}$B$_2$ and Fe$_2$Al$_{1.1}$B$_2$Ge$_{0.05}$Ga$_{0.05}$ through one-step reactive hot-pressing, resulting in dense, bulk, near single-phase MAB samples for magnetocaloric application. Hot-pressing is not only a much faster, processing cost reducing and more scalable synthesis route compared to conventional processes, but it can also directly produce net-shaped heat exchanger for magnetocaloric devices [60,61]. In this context, the adiabatic temperature change $\Delta T_{ad}$ for Fe$_2$AlB$_2$-type compounds is evaluated for the first time through direct measurements. Therefore, our work facilitates the pathway to low-criticality, Fe-based MAB phases for energy-efficient magnetic refrigeration around room temperature.

## 2. Experimental methods

### 2.1. Synthesis

The purities and sources of the powders used herein are listed in Table 1. Precursor powders were mixed and ball milled in a polyethylene jar for 24 h using ceria-stabilized zirconia milling balls with a diameter of 13 – 25 mm, a 2:1 ball to powder weight ratio and a rotation speed of 80 rpm. The powders were loaded in a graphite die and placed in a vacuum hot-press. To prevent reaction between the die wall and the powders, graphite foil is wrapped around the inner die walls as well as above and below the powders inserted in the die. Boron nitride (BN) was applied to the graphite die walls prior to layering with graphite foil to prevent sticking.

The samples were heated to 1473 K at a rate of 400 Kh$^{-1}$ and soaked for 4 h at 35 MPa load under low vacuum (< 20 Pa). The load was applied during the soaking stage after which the pressure was gradually released and the hot-press passively cooled to room temperature. Any bonded graphite or carbide layers were removed using diamond grinding pads prior to subsequent characterization.

For comparison with conventional processing methods, two $Fe_2Al_{1.1}B_2Ge_{0.05}Ga_{0.05}$ samples were synthesized by arc-melting under argon atmosphere, the technique which is commonly used for the production of MAB phases [47,49–54]. For the arc-molten samples, two different annealing conditions were used to homogenize the Ge and Ga distribution and to dissolve precipitates from the grain-boundaries into the MAB phase. The arc-molten samples were annealed at 1313 K for 72 h (ARC1) and at 1623 K for 1 h (ARC2).

**Table 1:** Specifications for reagent powders used for sample synthesis.

| Reagent | Supplier | Purity (metals basis) | Particle size [µm] |
|---|---|---|---|
| Fe | Alfa Aesar | 99.5%, reduced | 6-10 |
| Al | Alfa Aesar | 99.5% | < 44 |
| B, amorphous | US Nano | 99% | 1-2 nm |
| Ga | GALLANT | 99.9% | Ingot |
| Ge | Alfa Aesar | 99.999% | < 44 |

### 2.2. Characterization

The phase-purity and crystal structure have been analyzed with powder X-ray diffraction (XRD). The bulk samples have been milled to powder with a particle size < 80 µm. Room temperature XRD has been carried out with a *Stoe* Stadi P diffractometer in transmission geometry using Mo $K_{\alpha 1}$ radiation. Temperature-dependent powder XRD has been performed with a purpose-built diffractometer using Mo $K_\alpha$ radiation. A detailed description of the device is given by T. Faske *et al.* [62]. Standard silicon *NIST* SRM 640d powder has been used as reference in this setup. Structural analysis has been carried out with Rietveld-refinement using the *FullProf* software package [63] and for crystal structure visualization *VESTA* [64] was used.

Micrographs were obtained with a *Zeiss* Axio Imager.D2m optical microscope. Backscatter electron (BSE) and secondary electron (SE) imaging was performed using the scanning electron microscopes (SEM) *JOEL* JSM7600F and *Tescan* Vega3. Elemental compositions and elemental maps were obtained with the energy-dispersive X-ray spectroscopes (EDX) *Oxford* X-Max and *EDAX* Octane Plus. Impurity fractions were estimated by image contrast analysis using *ImageJ* software package [65].

Magnetic measurements have been performed with a *Quantum Design* physical property measurement system (PPMS-14T) and magnetic property measurement system (MPMS3). Isofield measurements have been carried out with a heating and cooling rate of 2 Kmin$^{-1}$. Isothermal measurements have been performed using a magnetic field ramp rate of 5 mTs$^{-1}$.

The isothermal entropy change $\Delta s_T(T, H)$ has been calculated based on isothermal magnetic measurements, using the Maxwell-relation [66,67]

$$\Delta s_T(T, H) = \mu_o \int_0^H \left(\frac{\partial M}{\partial T}\right)_H dH \quad (1)$$

with the permeability of free space $\mu_0$, magnetic field $H$, magnetization $M$ and temperature $T$.

The adiabatic temperature change $\Delta T_{ad}$ has been measured with a purpose-built device using a differential type T thermocouple and ice-water as temperature reference. The sinusoidal magnetic field profile has been applied with a maximum amplitude of 1.93 T and an average ramp rate of 0.4 Ts$^{-1}$ using a rotating double Halbach array. A detailed description of the device is given by J. Liu *et al.* [14].

## 3. Results and discussion

### 3.1. Structure and composition

Room temperature XRD patterns confirm the MAB phase (*Cmmm* space group, see Figure 1 (a)) being the majority phase of hot-pressed Fe$_2$Al$_{1.15}$B$_2$ and Fe$_2$Al$_{1.1}$B$_2$Ge$_{0.05}$Ga$_{0.05}$ as well as arc-molten ARC1 and ARC2 (see Figure 1 (b)). For ternary Fe$_2$Al$_{1.15}$B$_2$, near single-phase quality is obtained. FeB impurity (*Pnma* space group) reflexes are detectable for Fe$_2$Al$_{1.1}$B$_2$Ge$_{0.05}$Ga$_{0.05}$, ARC1 and ARC2. The highest phase fraction of FeB is present in ARC2 (23.2 wt%). The signal of any additional secondary phases is below the detection limit.

The MAB phase room temperature unit cell parameters for Fe$_2$Al$_{1.15}$B$_2$ are $a = 2.924$ Å, $b = 11.040$ Å, and $c = 2.873$ Å (see Table 2), which are in good agreement with the literature [46,48,49,52,56–58,68]. The addition of Ge and Ga alters the unit cell parameter to $a = 2.928$ Å, $b = 11.036$ Å, and $c = 2.874$ Å. The increase in $a$ and $c$ is in agreement with

R. Barua et al. [48], however, the decrease in $b$ contradicts it. ARC1 shows very similar parameters compared to the hot-pressed $Fe_2Al_{1.1}B_2Ge_{0.05}Ga_{0.05}$, however, ARC2 shows unit cell parameters more similar to $Fe_2Al_{1.15}B_2$. The FeB impurity phase shows unit cell parameters of approximately $a = 5.51$ Å, $b = 2.94$ Å, $c = 4.06$ Å in all samples, which is in agreement with references 69,70.

Temperature-dependent XRD (see Figure 1 (c)) shows the characteristic anisotropic thermal expansion of $Fe_2Al_{1.15}B_2$ with decreasing unit cell parameters $a$ and $b$ but an increasing parameter $c$ during cooling at temperatures $T < T_C$, which is in good agreement with references 56,57. The corresponding volume change (see Figure 1 (c) inset) shows three distinct temperature regions in which the coefficient of thermal expansion is constant. In the high-temperature region ($T > 290$ K), the coefficient of thermal expansion is highest with a value of $1.12 \cdot 10^{-5}$ K$^{-1}$ in accordance with reference 68. In the low-temperature region (100 K $< T < 290$ K), the coefficient of thermal expansion measures only $4.75 \cdot 10^{-6}$ K$^{-1}$, which is congruent with reference 57 and remarkably small, indicating the potential of MAB phases for near-zero thermal expansion applications. At cryogenic temperatures ($T < 100$ K), the volume is approximately constant.

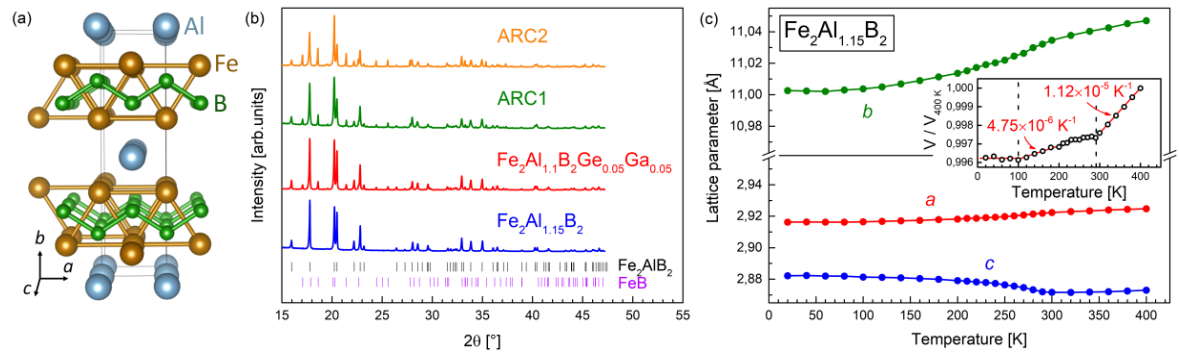

**Figure 1:** (a) Crystal structure illustration of $Fe_2AlB_2$ (*Cmmm* space group). (b) Room temperature powder XRD patterns of $Fe_2Al_{1.15}B_2$, $Fe_2Al_{1.1}B_2Ge_{0.05}Ga_{0.05}$, ARC1 and ARC2. (c) Temperature-dependent MAB phase unit cell parameter of $Fe_2Al_{1.15}B_2$. The inset shows the temperature-induced volume change and the coefficients of thermal expansion.

**Table 2:** MAB phase unit cell parameter and phase distribution at room temperature determined with XRD.

| Sample | Unit cell parameter | | | Phase fractions | |
|---|---|---|---|---|---|
| | $a$ [Å] | $b$ [Å] | $c$ [Å] | MAB [wt%] | FeB [wt%] |
| $Fe_2Al_{1.15}B_2$ | 2.9240 | 11.0402 | 2.8729 | 99.5 | 0.5 |
| $Fe_2Al_{1.1}B_2Ge_{0.05}Ga_{0.05}$ | 2.9283 | 11.0359 | 2.8745 | 90.1 | 9.9 |
| ARC1 | 2.9289 | 11.0338 | 2.8759 | 88.4 | 11.6 |
| ARC2 | 2.9321 | 11.0405 | 2.8712 | 76.8 | 23.2 |

Complementing the XRD results, SEM micrographs and EDX maps of all four samples are shown in Figure 2. The phase distribution determined with SEM image analysis is summarized in Table 3. The hot-pressed $Fe_2Al_{1.15}B_2$ (see Figure 2 (a-e)) is fully dense and 95.8% pure with 4.2% $Al_2O_3$ impurity most probably due to aluminothermic reduction of native oxides in metal precursor powders. Notably, hot-pressed highly dense $Fe_2Al_{1.1}B_2Ge_{0.05}Ga_{0.05}$ (see Figure 2 (f-l)) shows lower purity at 84.7% with the main impurity being binary FeB at 10.3% and $Al_2O_3$ at 4.8%. We also note the segregation of Ge-Ga-rich phases at the grain boundaries (see Figure 2 (k,l)). Therefore, Ge and Ga form precipitates as well as being incorporated into the MAB structure, as also observed by Barua *et al.* [48]. In both hot-pressed samples, the MAB phase consists of equiaxed grains with a grain size of approximately 25±10 µm (see supplementary material S1).

Both arc-molten samples show an increased porosity with respect to the hot-pressed samples (see Figure 2 (b,g,n,u)). The reference annealing condition used by Barua *et al.* [48] (ARC1) does not significantly impact MAB phase purity (82.7%) and FeB remains the main impurity phase at 8.5% (see Figure 2 (m-s)). The annealing temperature of 1623 K for 1 h (ARC2) leads to a significant reduction of the MAB phase purity to only 72.3%, primarily due to an increase in binary FeB impurities to 15.6% (see Figure 2 (t-z)) and Al-B-rich impurities with 1.2% (see supplementary material S2). Ge-Ga-rich intermetallic impurities persist as precipitates in ARC1 and ARC2 with 0.3% and 2.5%, respectively (see Figure 2 (r,s,y,z) and supplementary material S2). Our determination of decomposition products in ARC2 is in good agreement with the estimated thermal decomposition temperature of $Fe_2AlB_2$ at approximately 1509 K under vacuum [71].

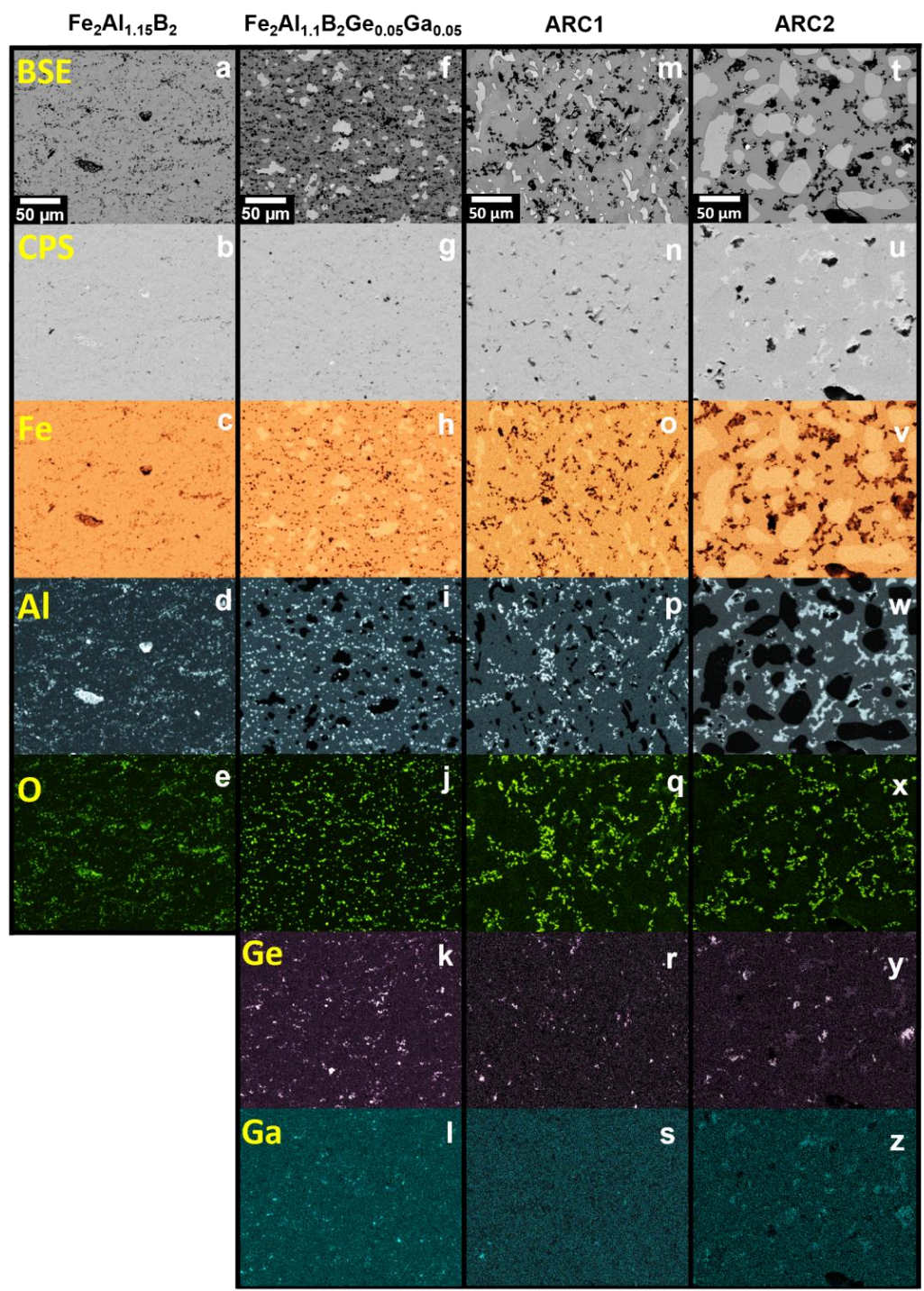

**Figure 2:** SEM-BSE images and qualitative element-specific EDX maps with X-ray counts per second (CPS) images of (a-e) $Fe_2Al_{1.15}B_2$, (f-l) $Fe_2Al_{1.1}B_2Ge_{0.05}Ga_{0.05}$, (m-s) ARC1 and (t-z) ARC2. The CPS images represent the statistics of the EDX maps and qualitatively show the porosity of each sample.

The Fe:Al ratio of the MAB phases is equal to 1.78, 1.89, 1.88 and 1.82 for $Fe_2Al_{1.15}B_2$, $Fe_2Al_{1.1}B_2Ge_{0.05}Ga_{0.05}$, ARC1 and ARC2, respectively. It should be noted that the absolute Fe:Al ratios are subjected to systematic errors associated with the composition determination by EDX, however, the relative composition differences between the MAB phases, which have been analyzed under identical conditions with the same system, are reliable.

**Table 3:** Phase distribution of MAB samples determined with SEM.

| Sample | Phase fractions | | | | |
|---|---|---|---|---|---|
| | MAB [%] | FeB [%] | $Al_2O_3$ [%] | Al-B-rich [%] | Ge-Ga-rich [%] |
| $Fe_2Al_{1.15}B_2$ | 95.8 | - | 4.2 | - | - |
| $Fe_2Al_{1.1}B_2Ge_{0.05}Ga_{0.05}$ | 84.7 | 10.3 | 4.8 | - | 0.2 |
| **ARC1** | 82.7 | 8.5 | 8.5 | - | 0.3 |
| **ARC2** | 72.3 | 15.6 | 8.5 | 1.2 | 2.5 |

### 3.2. Magnetic properties

Isofield magnetization curves in 1 T applied magnetic field of $Fe_2Al_{1.15}B_2$, $Fe_2Al_{1.1}B_2Ge_{0.05}Ga_{0.05}$, ARC1 and ARC2 are shown in Figure 3 (a). All samples show a transition from high-temperature paramagnetic to low-temperature ferromagnetic phase. The transition occurs at the Curie temperature $T_C$, which has been determined based on the extremum of $dMdT^{-1}$ in a magnetic field of 0.01 T (see supplementary material S3) and the linear extrapolation of the inverse magnetic susceptibility $\chi^{-1}(T)$ of the paramagnetic phase to $\chi^{-1}(T = T_C) = 0$ (see Figure 3 (a), right axis). To minimize the influence of secondary phases, the magnetic susceptibility $\chi$ has been obtained by linear regression of the isothermal magnetization curves in the magnetic field interval from 2.25 T to 2.75 T. Due to the non-linearity of $\chi^{-1}$ in the case of ARC2, the data is not used for further analysis. A significant increase of $T_C$ can be observed due to the Ge and Ga addition. Fe$_2$Al$_{1.15}$B$_2$, Fe$_2$Al$_{1.1}$B$_2$Ge$_{0.05}$Ga$_{0.05}$, ARC1 and ARC2 show a $T_C$ of approximately 284, 305, 303 and 299 K, respectively.

The ferromagnetic background in the paramagnetic state at $T = 390$ K of hot-pressed Fe$_2$Al$_{1.1}$B$_2$Ge$_{0.05}$Ga$_{0.05}$, ARC1 and ARC2 (see Figure 3 (a), inset) originates from the ferromagnetic FeB impurity phase ($T_C = 598$ K [72]) and scales with its phase fraction, consistent with the powder XRD (see Table 2) and SEM (see Table 3) results. The effective magnetic moment $\mu_{eff}$ of the paramagnetic phase has been determined by extracting the Curie constant from the linear regression of $\chi^{-1}(T)$ and amounts to 1.50, 1.64 and 1.64 µ$_B$ / Fe atom for Fe$_2$Al$_{1.15}$B$_2$, Fe$_2$Al$_{1.1}$B$_2$Ge$_{0.05}$Ga$_{0.05}$ and ARC1. The spontaneous magnetization $m_S$ has been extracted at 10 K (see supplementary material S4) and is corrected for the ferromagnetic contribution of the secondary FeB phase, using the phase fraction determined by XRD and the literature $m_S$ value of 77 Am$^2$kg$^{-1}$ for FeB [73]. Fe$_2$Al$_{1.15}$B$_2$, Fe$_2$Al$_{1.1}$B$_2$Ge$_{0.05}$Ga$_{0.05}$, ARC1 and ARC2 show a spontaneous magnetization of 76.3, 81.8, 82.4 and 73.5 Am$^2$kg$^{-1}$, i.e. 1.12, 1.25, 1.26, 1.12 µ$_B$ / Fe atom, respectively. The spontaneous magnetization is significantly smaller compared to the effective magnetic moment of the paramagnetic phase, leading to a Rhode-Wohlfarth ratio of $\mu_{eff} m_S^{-1} > 1$, which shows that the magnetism is of itinerant nature in the MAB samples, as observed in references 46–48.

The values of $T_C$, $m_S$ and $\mu_{eff}$ (see Table 4) are in good agreement for Fe$_2$Al$_{1.15}$B$_2$ with references 44–55 and for hot-pressed Fe$_2$Al$_{1.1}$B$_2$Ge$_{0.05}$Ga$_{0.05}$ and ARC1 with Barua *et al.* [48]. Comparing arc-molten with hot-pressed Fe$_2$Al$_{1.1}$B$_2$Ge$_{0.05}$Ga$_{0.05}$ reveals that the extended annealing time of ARC1 results in very similar properties as the hot-pressed sample, as expected from the XRD and SEM results. However, due to its thermal decomposition, ARC2 shows reduced $T_C$ and $m_S$ values as well as a less sharp phase transition (see Figure 3 and supplementary material S2), indicating a chemically inhomogeneous MAB phase, which results in the observable broad distribution of locally varying Curie temperatures. Comparing all four samples shows that $T_C$ increases with increasing Fe:Al ratio, revealing a general trend in this material system as previous experimental and theoretical studies have shown a decreasing $T_C$ upon partial Fe-substitution by other 3d transition metals such as Cr, Mn, Ni and Co [55,74,75].

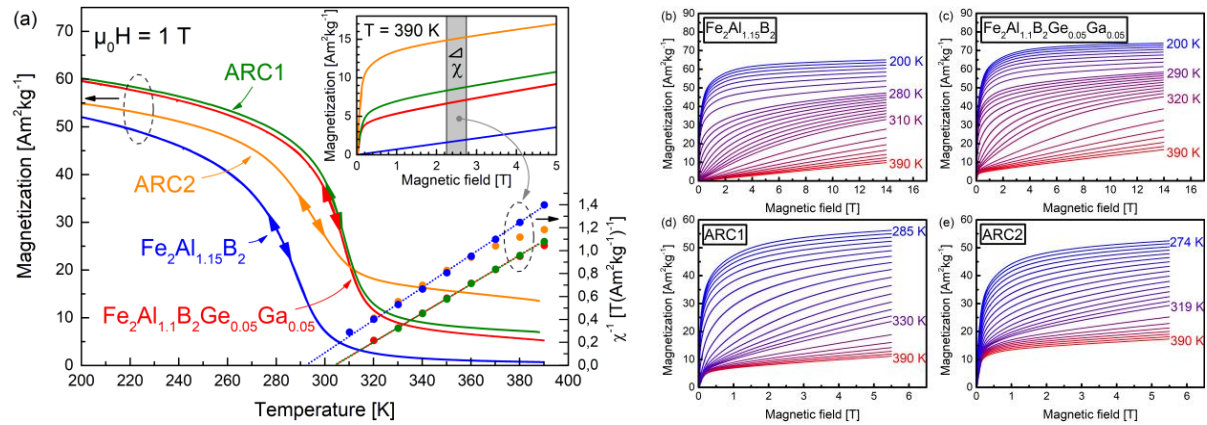

**Figure 3:** (a) Isofield magnetization curves in 1 T applied magnetic field (left axis) and inverse magnetic susceptibility $\chi^{-1}$ (right axis) of $Fe_2Al_{1.15}B_2$, $Fe_2Al_{1.1}B_2Ge_{0.05}Ga_{0.05}$, ARC1 and ARC2. The inset shows the isothermal magnetization curves at 390 K. Isothermal magnetization curves of (b) $Fe_2Al_{1.15}B_2$, (c) $Fe_2Al_{1.1}B_2Ge_{0.05}Ga_{0.05}$, (d) ARC1 and (e) ARC2 used to calculate the isothermal entropy change $\Delta s_T$ as well as magnetic susceptibility $\chi$, with temperature steps of 3 K near $T_C$ and 10 K far below and above $T_C$. The temperatures at which the temperature step size changes are indicated.

Based on the $dMdT^{-1}$ values observable in isofield and isothermal magnetization curves shown in Figure 3, similarly large isothermal entropy changes can be expected in hot-pressed $Fe_2Al_{1.1}B_2Ge_{0.05}Ga_{0.05}$ and ARC1 compared to hot-pressed $Fe_2Al_{1.15}B_2$ and ARC2 (see equation 1).

### 3.3. Magnetocaloric properties

The magnetocaloric effect has been characterized based on the isothermal entropy change $\Delta s_T$ and adiabatic temperature change $\Delta T_{ad}$ for $Fe_2Al_{1.15}B_2$, $Fe_2Al_{1.1}B_2Ge_{0.05}Ga_{0.05}$, ARC1 and ARC2 and is summarized in Table 4. The isothermal entropy change has been calculated (see equation 1) based on isothermal magnetization curves (see Figure 3 (b-e)) and is shown for magnetic field changes up to 13.9 T in Figure 4.

$Fe_2Al_{1.15}B_2$, $Fe_2Al_{1.1}B_2Ge_{0.05}Ga_{0.05}$, ARC1 and ARC2 show maximum $|\Delta s_T|$ values of 2.5, 3.5, 3.4 and 1.5 J(kgK)$^{-1}$ in magnetic field changes of 2 T and 5.2, 6.7, 6.7 and 3.6 J(kgK)$^{-1}$ in magnetic field changes of 5 T around $T_C$, respectively. Thereby, $\Delta s_{T,max}^{2T}$ improves by 40% by introducing small quantities of Ge and Ga to the MAB phase. No significant difference between hot-pressed $Fe_2Al_{1.1}B_2Ge_{0.05}Ga_{0.05}$ and ARC1 can be detected. However, ARC2 shows not only a significant reduction in $\Delta s_T$, but also a broader $\Delta s_T(T)$ distribution compared to the other samples. This is quantified by the full width at half maximum (FWHM) $\Delta s_{T,FWHM}^{2T}$ value of 27.6, 15.7, 16.4 and 33.4 K for $Fe_2Al_{1.15}B_2$, $Fe_2Al_{1.1}B_2Ge_{0.05}Ga_{0.05}$, ARC1 and ARC2, respectively. In general, $\Delta s_T$ agrees very well to the expectations formulated in the context of the isofield and isothermal magnetization measurements (see Figure 3).

For all compounds, the isothermal entropy change at $T_C$ scales according to $\Delta s_T \propto H^n$ (see Figure 4 insets), obeying the scaling law applicable to all second-order phase transitions in magnetocaloric materials [76]. The exponent $n$ determined by non-linear fitting of $\Delta s_T$ for $Fe_2Al_{1.15}B_2$, $Fe_2Al_{1.1}B_2Ge_{0.05}Ga_{0.05}$, ARC1 and ARC2 at $T \approx T_C$ is equal to 0.69, 0.62, 0.79 and 0.95, which is in good agreement with $n = \frac{d \ln |\Delta s_T|}{d \ln H}$ resulting in 0.77, 0.68, 0.85 and 0.99, respectively, underlining the consistency of both determination methods. A more detailed

analysis and discussion of $n$ is omitted due to the multiphase character of the samples, however, the exponent is still very valuable to accurately describe $\Delta s_T(H)$ at $T \approx T_C$ for each sample.

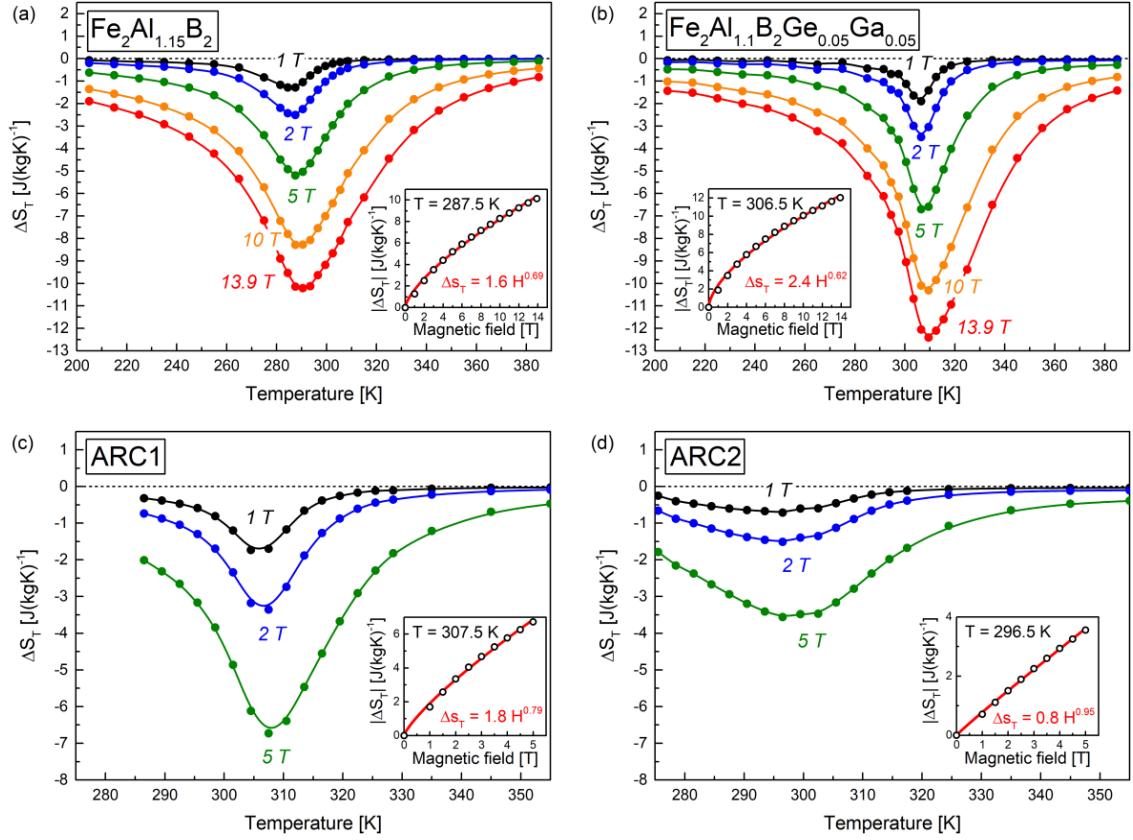

**Figure 4:** Isothermal entropy change $\Delta s_T$ of (a) $Fe_2Al_{1.15}B_2$, (b) $Fe_2Al_{1.1}B_2Ge_{0.05}Ga_{0.05}$, (c) ARC1 and (d) ARC2. The insets show the magnetic field dependencies of $\Delta s_T$ at $T \approx T_C$.

The directly measured adiabatic temperature change $\Delta T_{ad}$ for a magnetic field change of 1.93 T is shown in Figure 5. $Fe_2Al_{1.15}B_2$ shows a maximum in $\Delta T_{ad}$ of 0.9 K at 286 K. The adiabatic temperature change is improved by 22% to 1.1 K at 304 K for $Fe_2Al_{1.1}B_2Ge_{0.05}Ga_{0.05}$. The influence of processing parameters on the $\Delta T_{ad}$ can be seen in Figure 5 (b). Arc melting followed by annealing at 1313 K for 72 h (ARC1) only slightly increased the $\Delta T_{ad}$ value to 1.2 K at 304 K with respect to the hot-pressed $Fe_2Al_{1.1}B_2Ge_{0.05}Ga_{0.05}$ sample. However, annealing at 1623 K for 1 h (ARC2) strongly deteriorated the performance, showing only a $\Delta T_{ad}$ of 0.6 K at 295 K and a very broad $\Delta T_{ad}(T)$ distribution. The $\Delta T_{ad,FWHM}$ shows very similar values compared with $\Delta s_{T,FWHM}^{2T}$, namely 23.2, 16.7, 15.6 and 36.1 K for $Fe_2Al_{1.15}B_2$, $Fe_2Al_{1.1}B_2Ge_{0.05}Ga_{0.05}$, ARC1 and ARC2, respectively. Therefore, the $\Delta T_{ad}$ results fit very well to the magnetic measurements and $\Delta s_T$ results shown in Figures 3 and 4, respectively. All samples show a fully reversible magnetic field dependence of $\Delta T_{ad}$ and the absence of thermal and magnetic hysteresis underlines the second-order nature of the phase transition and is consistent with Figure 3. This is supported by Arrott plots (see supplementary material S5), showing a second-order phase transition character based on the Banerjee criterion [77], supporting previous assessments in the literature [45,50,51,53,56]. It should be noted that Lejeune *et al.* [78] considered the phase transition of $Fe_2Al_{1.1}B_2Ga_{0.1}$ to be of "borderline first-order-type transformation character" based on heat capacity and lab-scale XRD measurements. However, to unambiguously identify the first-order character of the phase transition by observing the coexistence of both phases, high-resolution synchrotron XRD should be performed in the future, comparable to the study by Oey *et al.* [56] for $Fe_2AlB_2$.

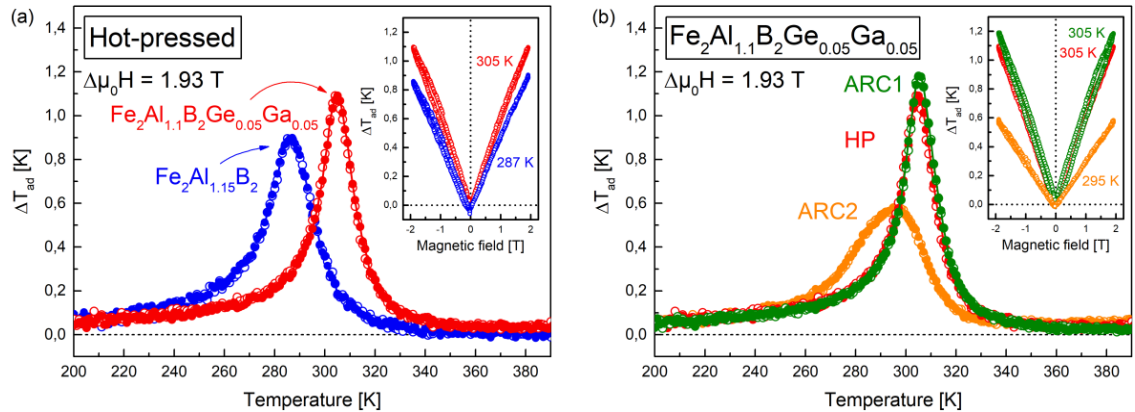

**Figure 5:** (a) Adiabatic temperature change $\Delta T_{ad}$ of hot-pressed $Fe_2Al_{1.15}B_2$ and $Fe_2Al_{1.1}B_2Ge_{0.05}Ga_{0.05}$ for a magnetic field change of 1.93 T during cooling (open circles) and heating protocol (solid circles). (b) Adiabatic temperature change of hot-pressed (HP) and arc-molten (ARC1, ARC2) $Fe_2Al_{1.1}B_2Ge_{0.05}Ga_{0.05}$ samples in a magnetic field change of 1.93 T. The insets show the magnetic field dependencies of $\Delta T_{ad}$ at $T \approx T_C$.

Comparing the magnetocaloric effect with the literature shows that the isothermal entropy change values for hot-pressed $Fe_2Al_{1.15}B_2$ in magnetic field changes of 2 and 5 T are in agreement with reported values [45,53], higher compared to reference 79 and lower than single-crystalline [46] and pressure-less argon sintered polycrystalline $Fe_2AlB_2$ [80]. The adiabatic temperature change $\Delta T_{ad}$ values for $Fe_2Al_{1.15}B_2$ measured herein at 1.93 T are somewhat lower compared to samples that were produced by arc-melting in references 48,49. The reason for this discrepancy is not entirely understood, however, Tan *et al.* [49] used an indirect method to determine $\Delta T_{ad}$ and report the use of Ga flux and an acid treatment during synthesis, which can affect magnetocaloric performance.

**Table 4:** Magnetic and magnetocaloric properties of the MAB samples compared to the literature with maximum isothermal entropy changes $\Delta s_{T,max}^{2T}$ ($\mu_0 \Delta H = 2$ T) and $\Delta s_{T,max}^{5T}$ ($\mu_0 \Delta H = 5$ T), maximum adiabatic temperature change $\Delta T_{ad,max}^{1.93T}$ ($\mu_0 \Delta H = 1.93$ T), spontaneous magnetization $m_S$ at 10 K, Curie temperature $T_C$ and effective magnetic moment $\mu_{eff}$.

| Sample | $\Delta s_{T,max}^{2T}$ [J(kgK)$^{-1}$] | $\Delta s_{T,max}^{5T}$ [J(kgK)$^{-1}$] | $\Delta T_{ad,max}^{1.93T}$ [K] | $m_S$ [Am$^2$kg$^{-1}$] | $T_C$ [K] | $\mu_{eff}$ [$\mu_B$ / Fe] | Ref. |
|---|---|---|---|---|---|---|---|
| **Fe$_2$Al$_{1.15}$B$_2$** | -2.5 | -5.2 | 0.9 | 76.3$^a$, 76.3$^b$ | 284$^c$, 292$^d$ | 1.50 | This work |
| **Fe$_2$Al$_{1.1}$B$_2$Ge$_{0.05}$Ga$_{0.05}$** | -3.5 | -6.7 | 1.1 | 81.3$^a$, 81.8$^b$ | 305$^{c,d}$ | 1.64 | This work |
| **ARC1** | -3.4 | -6.7 | 1.2 | 81.8$^a$, 82.4$^b$ | 303$^c$, 304$^d$ | 1.64 | This work |
| **ARC2** | -1.5 | -3.6 | 0.6 | 74.3$^a$, 73.5$^b$ | 299$^c$ | - | This work |
| **Fe$_2$AlB$_2$** | -2.7 | - | 1 ($\mu_0\Delta H=2$ T) | 78 (T=50 K) | 272 | 1.61 | 48 |
| **Fe$_2$Al$_{1.1}$B$_2$Ge$_{0.05}$Ga$_{0.05}$** | -6.5 | - | 2.2 ($\mu_0\Delta H=2$ T) | 85 (T=50 K) | 294 | 1.62 | 48 |

$^a$ Uncorrected
$^b$ Corrected for ferromagnetic contribution of FeB
$^c$ Determined with $(dMdT^{-1})_{max}$ in 0.01 T
$^d$ Determined with linear extrapolation of $\chi^{-1}(T > T_C)$ to $\chi^{-1}(T = T_C) = 0$

According to Barua *et al.* [48], the increase in the Fe:Al ratio, due to antisite occupancy of Fe into the Al lattice, is likely the reason for the improved magnetocaloric performance of $Fe_2Al_{1.1}B_2Ge_{0.05}Ga_{0.05}$ compared to $Fe_2Al_{1.15}B_2$. However, there is a discrepancy between the magnetocaloric properties of hot-pressed and arc-molten $Fe_2Al_{1.1}B_2Ge_{0.05}Ga_{0.05}$ sintered herein and those reported by Barua *et al*. This disparity could be due to the difference between conventional arc-melting and suction-casting, resulting in distinct microstructures in terms of grain size, secondary phases and texture. Especially the impurity phases, acting as magnetocaloric inactive material, are implicit in the different magnetocaloric properties of the hot-pressed and arc-molten Fe$_2$Al$_{1.1}$B$_2$Ge$_{0.05}$Ga$_{0.05}$ compared to Barua *et al.* [48].

Interestingly, the ARC1 annealing condition of 1313 K for 72 h, which reproduces the heat treatment used by Barua *et al.*, leads only to a small improvement in $\Delta T_{ad}$ compared to reactive hot-pressing of Fe$_2$Al$_{1.1}$B$_2$Ge$_{0.05}$Ga$_{0.05}$. This indicates, in combination with FeB being the major secondary phase in hot-pressed Fe$_2$Al$_{1.1}$B$_2$Ge$_{0.05}$Ga$_{0.05}$, that an increase of Al excess during synthesis might bypass this limitation, leading to an increased purity of hot-pressed Ge and Ga doped Fe$_2$AlB$_2$.

To evaluate the magnetocaloric performance of hot-pressed Fe$_2$Al$_{1.15}$B$_2$ and Fe$_2$Al$_{1.1}$B$_2$Ge$_{0.05}$Ga$_{0.05}$, their properties are compared to other metallic magnetocaloric materials showing second-order phase transitions close to room temperature (see Table 5 and Figure 6). The comparison includes the maximum adiabatic temperature change $\Delta T_{ad,max}^{2T}$ and isothermal entropy change $\Delta s_{T,max}^{2T}$ for field changes of approximately 2 T, the estimated material cost $c_{mat}$, the resulting material cost specific maximum isothermal entropy change $\Delta s_{T,max}^{2T} c_{mat}^{-1}$ as well as the material criticality and Curie temperature $T_C$. The material cost is estimated based on the sum of averaged element prices extracted from well-known and widely-used databases [81–83]. The price for B and Gd is estimated based on B$_2$O$_3$ [84] and Gd$_2$O$_3$ [81–83], considering a mass conversion factor of 0.31 and 0.87, respectively. Processing costs are neglected in this estimation, which thereby represents the lower bound of total costs associated with the magnetocaloric material. The material criticality is assessed on a scale from 1 (low) to 6 (high), following the procedure described by T. Gottschall *et al.* [8]. The criticality score of each material is based on the geological availability and recyclability of its constituent elements as well as the sustainability of their extraction processes and their geopolitical importance [36–40]. The comparison reveals that the herein reported MAB phases show a very promising compromise of $\Delta s_{T,max}^{2T} c_{mat}^{-1}$, criticality and magnetocaloric performance, demonstrating the potential of Fe$_2$AlB$_2$-type MAB phases for low-cost and low-criticality magnetocaloric applications. Therefore, future studies should focus on the challenging task to improve the magnetocaloric properties by increasing $m_S$ and $dMdT^{-1}(T=T_C)$ while keeping $T_C$ at room temperature, using only low-cost and low-criticality elements. Recently, Chen *et al.* [85] presented a theory-driven study about the design of MAB phases for energy applications, observing a positive correlation between isothermal entropy change and lattice deformation during the phase transition, suggesting on this basis several novel MAB phases for magnetic cooling applications. However, future experimental follow-up studies are required to determine whether these phases can be synthesized as well as the $T_C$ and magnetocaloric effect of the respective phase transitions.

**Table 5:** Comparison of metallic magnetocaloric materials showing second-order phase transitions with $T_C$ close to room temperature. Hot-pressed Fe$_2$Al$_{1.15}$B$_2$ and Fe$_2$Al$_{1.1}$B$_2$Ge$_{0.05}$Ga$_{0.05}$ are compared with Gd, Gd$_{98.7}$Y$_{1.3}$, NiFeCoCrPd$_{0.5}$, Ni$_2$Mn$_{1.4}$In$_{0.6}$ and Ni$_{33}$Co$_{17}$Mn$_{30}$Ti$_{20}$ in terms of the estimated material cost $c_{mat}$, maximum isothermal entropy change $\Delta s_{T,max}^{2T}$, resulting material cost specific maximum isothermal entropy change $\Delta s_{T,max}^{2T} c_{mat}^{-1}$, maximum adiabatic temperature change $\Delta T_{ad,max}^{2T}$ as well as criticality and Curie temperature $T_C$.

| Second order phase transition | $c_{mat}$ [\$/kg] | $\|\Delta s_{T,max}^{2T}\|$ [J(kgK)$^{-1}$] | $\|\Delta s_{T,max}^{2T}\|c_{mat}^{-1}$ [J(k\$K)$^{-1}$] | $\Delta T_{ad,max}^{2T}$ [K] | Criticality | $T_C$ [K] | Reference |
|---|---|---|---|---|---|---|---|
| **Fe$_2$Al$_{1.15}$B$_2$** | 2 | 2.5 | 1250 | 0.9 (μ$_0$ΔH=1.93 T) | 2 | 284 | This work |
| **Fe$_2$Al$_{1.1}$B$_2$Ge$_{0.05}$Ga$_{0.05}$** | 48 | 3.5 | 73 | 1.1 (μ$_0$ΔH=1.93 T) | 3 | 305 | This work |
| **Gd** | 63 | 5.2 | 83 | 4.7 | 5 [8] | 292 | 8 |
| **Gd$_{98.7}$Y$_{1.3}$** | 63 | 5.0 | 79 | 4.1 | 5 [8] | 284 | 8 |
| **NiFeCoCrPd$_{0.5}$** | 13571 | 0.8 (μ$_0$ΔH=5 T) | 0.06 | - | 6 | 300 | 29 |
| **Ni$_2$Mn$_{1.4}$In$_{0.6}$** | 65 | 3.3 | 51 | 2 | 4 | 316 | 30 |
| **Ni$_{33}$Co$_{17}$Mn$_{30}$Ti$_{20}$** | 21 | 1.1 | 52 | 0.6[a] | 5 | 275 | 31 |

[a] Estimated based on $\Delta T_{ad,max}^{1T} = 0.3$ K

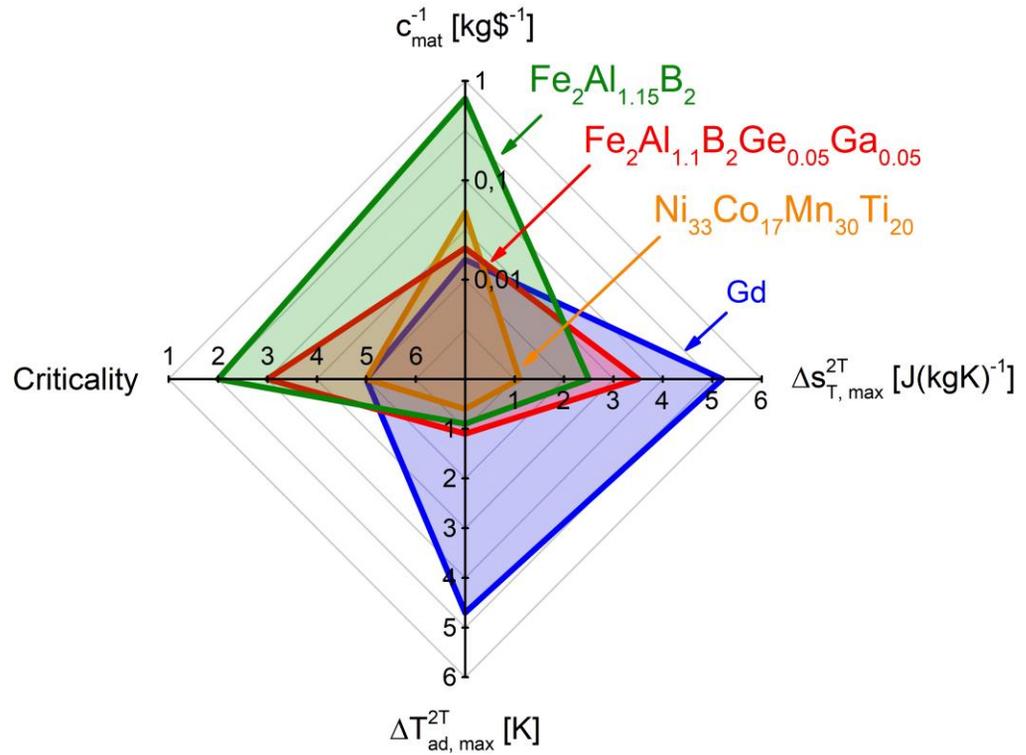

**Figure 6:** Comparison of metallic magnetocaloric materials showing second-order phase transitions with $T_C$ close to room temperature. Hot-pressed $Fe_2Al_{1.15}B_2$ and $Fe_2Al_{1.1}B_2Ge_{0.05}Ga_{0.05}$ are compared with Gd and $Ni_{33}Co_{17}Mn_{30}Ti_{20}$ in terms of material cost $c_{mat}$, criticality, maximum isothermal entropy change $\Delta s_{T,max}^{2T}$ and adiabatic temperature change $\Delta T_{ad,max}^{2T}$ in magnetic field changes of 2 T. Note the logarithmic scale for $c_{mat}^{-1}$. More detailed information is given in Table 5.

## 4. Conclusion

Reactive single-step hot-pressing of Fe, Al, B, Ge and Ga at 1473 K and 35 MPa for 4 h produces fully dense, near single-phase, low-cost and low-criticality $Fe_2Al_{1.15}B_2$ and $Fe_2Al_{1.1}B_2Ge_{0.05}Ga_{0.05}$ MAB samples with favorable magnetocaloric properties based on their second-order magnetic transition around room temperature, substantially improving the annealing time compared to the 72 h annealing method of suction cast material extracted from the literature.

The magnetic as well as the magnetocaloric properties of hot-pressed $Fe_2Al_{1.15}B_2$ can be tailored and improved by the addition of Ge and Ga. Most importantly, the maximum isothermal entropy change $\Delta s_{T,max}$ of $Fe_2Al_{1.15}B_2$ in magnetic field changes of 2 T amounts to 2.5 J(kgK)$^{-1}$ and increases by the Ge and Ga addition to 3.1 J(kgK)$^{-1}$, respectively. The directly measured maximum adiabatic temperature change $\Delta T_{ad,max}$ is improved by the doping from 0.9 to 1.1 K in magnetic field changes of 1.93 T. As the Fe$_2$Al$_{1.1}$B$_2$Ge$_{0.05}$Ga$_{0.05}$ arc-molten reference sample shows very similar phase purity and magnetocaloric performance compared to the hot-pressed sample, we successfully demonstrate the substitution of the conventional processing route with hot-pressing and highlight the potential of hot-pressed Fe$_2$AlB$_2$-type MAB phases as low-cost and low-criticality heat exchanger material for magnetocaloric applications around room temperature.

**Supplementary material**

See supplementary material for additional optical microscopy and SEM-BSE images of hot-pressed Fe$_2$Al$_{1.15}$B$_2$, Fe$_2$Al$_{1.1}$B$_2$Ge$_{0.05}$Ga$_{0.05}$, and arc-molten Fe$_2$Al$_{1.1}$B$_2$Ge$_{0.05}$Ga$_{0.05}$ (ARC2) as well as normalized isofield magnetization curves in 0.01 T, isothermal magnetization curves at 10 K and Arrott plots of all samples.

**Acknowledgements**

We acknowledge financial support by the Deutsche Forschungsgemeinschaft (DFG) within the CRC/TRR 270 (Project-ID 405553726). The research was also supported by National Science Foundation (1729335). The authors would like to thank S. Kota for valuable contribution to the synthesis of the MAB phases.

**Data availability**

The data that support the findings of this study are available from the corresponding author upon reasonable request.

**References**

[1] M. Isaac and D.P. van Vuuren, Energy Policy **37**, 507 (2009).

[2] United Nations, Department of Economic and Social Affairs, and Population Division, *World Population Prospects 2019: Highlights, United Nations* (2019).

[3] S. Qian, D. Nasuta, A. Rhoads, Y. Wang, Y. Geng, Y. Hwang, R. Radermacher, and I. Takeuchi, Int. J. Refrig. **62**, 177 (2016).

[4] D. Coulomb, V. Morlet, and L.J. Dupont, *35th Informatory Note on Refrigeration Technologies: The Impact of the Refrigeration Sector on Climate Change, International Institute of Refrigeration (IIR)* (2017).

[5] O. Gutfleisch, M.A. Willard, E. Brück, C.H. Chen, S.G. Sankar, and J.P. Liu, Adv. Mater. **23**, 821 (2011).

[6] V. Franco, J.S. Blázquez, J.J. Ipus, J.Y. Law, L.M. Moreno-Ramírez, and A. Conde, Prog. Mater. Sci. **93**, 112 (2018).

[7] X. Moya, S. Kar-Narayan, and N.D. Mathur, Nat. Mater. **13**, 439 (2014).

[8] T. Gottschall, K.P. Skokov, M. Fries, A. Taubel, I. Radulov, F. Scheibel, D. Benke, S. Riegg, and O. Gutfleisch, Adv. Energy Mater. **9**, 1901322 (2019).

[9] V. Chaudhary, X. Chen, and R.V. Ramanujan, Prog. Mater. Sci. **100**, 64 (2019).

[10] V.K. Pecharsky and K.A. Gschneidner, Jr., Phys. Rev. Lett. **78**, 4494 (1997).

[11] S.A. Nikitin, G. Myalikgulyev, A.M. Tishin, M.P. Annaorazov, K.A. Asatryan, and A.L. Tyurin, Phys. Lett. A **148**, 363 (1990).

[12] A. Chirkova, K.P. Skokov, L. Schultz, N.V. Baranov, O. Gutfleisch, and T.G. Woodcock, Acta Mater. **106**, 15 (2016).

[13] A. Taubel, B. Beckmann, L. Pfeuffer, N. Fortunato, F. Scheibel, S. Ener, T. Gottschall, K.P. Skokov, H. Zhang, and O. Gutfleisch, Acta Mater. **201**, 425 (2020).

[14] J. Liu, T. Gottschall, K.P. Skokov, J.D. Moore, and O. Gutfleisch, Nat. Mater. **11**, 620 (2012).

[15] T. Krenke, E. Duman, M. Acet, E.F. Wassermann, X. Moya, L. Mañosa, and A. Planes, Nat. Mater. **4**, 450 (2005).

[16] O. Gutfleisch, A. Yan, and K.-H. Müller, J. Appl. Phys. **97**, 10M305 (2005).

[17] A. Fujita, S. Fujieda, Y. Hasegawa, and K. Fukamichi, Phys. Rev. B **67**, 104416 (2003).

[18] J. Lyubina, K. Nenkov, L. Schultz, and O. Gutfleisch, Phys. Rev. Lett. **101**, 177203 (2008).

[19] O. Tegus, E. Brück, K.H.J. Buschow, and F.R. de Boer, Nature **415**, 150 (2002).

[20] M. Fries, L. Pfeuffer, E. Bruder, T. Gottschall, S. Ener, L.V.B. Diop, T. Gröb, K.P. Skokov, and O. Gutfleisch, Acta Mater. **132**, 222 (2017).

[21] V. Provenzano, A.J. Shapiro, and R.D. Shull, Nature **429**, 853 (2004).

[22] O. Gutfleisch, T. Gottschall, M. Fries, D. Benke, I. Radulov, K.P. Skokov, H. Wende, M. Gruner, M. Acet, P. Entel, and M. Farle, Philos. Trans. R. Soc. A Math. Phys. Eng. Sci. **374**, 20150308 (2016).

[23] L.F. Cohen, Phys. Status Solidi **255**, 1700317 (2018).

[24] B. Beckmann, D. Koch, L. Pfeuffer, T. Gottschall, A. Taubel, E. Adabifiroozjaei, O.N.

Miroshkina, S. Riegg, T. Niehoff, N.A. Kani, M.E. Gruner, L. Molina-Luna, K.P. Skokov, and O. Gutfleisch, Acta Mater. 118695 (2023).

[25] A. Waske, L. Giebeler, B. Weise, A. Funk, M. Hinterstein, M. Herklotz, K. Skokov, S. Fähler, O. Gutfleisch, and J. Eckert, Phys. Status Solidi - Rapid Res. Lett. **9**, 136 (2015).

[26] J. Liu, N. Scheerbaum, J. Lyubina, and O. Gutfleisch, Appl. Phys. Lett. **93**, 102512 (2008).

[27] V. Basso, C.P. Sasso, K.P. Skokov, O. Gutfleisch, and V. V. Khovaylo, Phys. Rev. B **85**, 014430 (2012).

[28] G. V. Brown, J. Appl. Phys. **47**, 3673 (1976).

[29] D.D. Belyea, M.S. Lucas, E. Michel, J. Horwath, and C.W. Miller, Sci. Rep. **5**, 15755 (2015).

[30] S. Singh, L. Caron, S.W. D'Souza, T. Fichtner, G. Porcari, S. Fabbrici, C. Shekhar, S. Chadov, M. Solzi, and C. Felser, Adv. Mater. **28**, 3321 (2016).

[31] F. Zhang, K. Westra, Q. Shen, I. Batashev, A. Kiecana, N. van Dijk, and E. Brück, J. Alloys Compd. **906**, 164337 (2022).

[32] A. El Boukili, N. Tahiri, E. Salmani, H. Ez-Zahraouy, M. Hamedoun, A. Benyoussef, M. Balli, and O. Mounkachi, Intermetallics **104**, 84 (2019).

[33] X. Zhang, B.T. Lejeune, R. Barua, R.W. McCallum, and L.H. Lewis, J. Alloys Compd. **823**, 153693 (2020).

[34] A. Kitanovski, J. Tušek, U. Tomc, U. Plaznik, M. Ožbolt, and A. Poredoš, *Magnetocaloric Energy Conversion* (Springer Cham, 2015).

[35] F. Scarpa, G. Tagliafico, and L.A. Tagliafico, Renew. Sustain. Energy Rev. **50**, 497 (2015).

[36] European Commission, Directorate-General for Internal Market, Industry,

Entrepreneurship and SMEs, G. Blengini, C. El Latunussa, U. Eynard, C. Torres De Matos, D. Wittmer, K. Georgitzikis, C. Pavel, S. Carrara, L. Mancini, M. Unguru, D. Blagoeva, F. Mathieux, and D. Pennington, *Study on the EU's List of Critical Raw Materials (2020) : Executive Summary* (Publications Office, 2020).

[37] R. Gauß, G. Homm, and O. Gutfleisch, J. Ind. Ecol. **21**, 1291 (2017).

[38] S. Glöser, L. Tercero Espinoza, C. Gandenberger, and M. Faulstich, Resour. Policy **44**, 35 (2015).

[39] K. Binnemans, P.T. Jones, T. Müller, and L. Yurramendi, J. Sustain. Metall. **4**, 126 (2018).

[40] J.P. Sykes, J.P. Wright, and A. Trench, Appl. Earth Sci. **125**, 3 (2016).

[41] S. Kota, M. Sokol, and M.W. Barsoum, Int. Mater. Rev. **65**, 226 (2020).

[42] W. Jeitschko, Acta Crystallogr. Sect. B Struct. Crystallogr. Cryst. Chem. **25**, 163 (1969).

[43] Y.B. Kuzma and N.F. Chaban, Izvest Akad Nauk SSSR Neorg Mater. **5**, 384 (1969).

[44] E.M. Levin, B.A. Jensen, R. Barua, B. Lejeune, A. Howard, R.W. McCallum, M.J. Kramer, and L.H. Lewis, Phys. Rev. Mater. **2**, 034403 (2018).

[45] Q. Du, G. Chen, W. Yang, Z. Song, M. Hua, H. Du, C. Wang, S. Liu, J. Han, Y. Zhang, and J. Yang, Jpn. J. Appl. Phys. **54**, 53003 (2015).

[46] T.N. Lamichhane, L. Xiang, Q. Lin, T. Pandey, D.S. Parker, T.H. Kim, L. Zhou, M.J. Kramer, S.L. Bud'Ko, and P.C. Canfield, Phys. Rev. Mater. **2**, 84408 (2018).

[47] M. ElMassalami, D.D.S. Oliveira, and H. Takeya, J. Magn. Magn. Mater. **323**, 2133 (2011).

[48] R. Barua, B.T. Lejeune, B.A. Jensen, L. Ke, R.W. McCallum, M.J. Kramer, and L.H. Lewis, J. Alloys Compd. **777**, 1030 (2019).

[49] X. Tan, P. Chai, C.M. Thompson, and M. Shatruk, J. Am. Chem. Soc. **135**, 9553 (2013).

[50] J. Cedervall, M.S. Andersson, T. Sarkar, E.K. Delczeg-Czirjak, L. Bergqvist, T.C. Hansen, P. Beran, P. Nordblad, and M. Sahlberg, J. Alloys Compd. **664**, 784 (2016).

[51] T. Ali, M.N. Khan, E. Ahmed, and A. Ali, Prog. Nat. Sci. Mater. Int. **27**, 251 (2017).

[52] R. Barua, B.T. Lejeune, L. Ke, G. Hadjipanayis, E.M. Levin, R.W. McCallum, M.J. Kramer, and L.H. Lewis, J. Alloys Compd. **745**, 505 (2018).

[53] J.W. Lee, M.S. Song, K.K. Cho, B.K. Cho, and C. Nam, J. Korean Phys. Soc. **73**, 1555 (2018).

[54] J. Cedervall, L. Häggström, T. Ericsson, and M. Sahlberg, Hyperfine Interact. **237**, 47 (2016).

[55] S. Hirt, F. Yuan, Y. Mozharivskyj, and H. Hillebrecht, Inorg. Chem. **55**, 9677 (2016).

[56] Y.M. Oey, J.D. Bocarsly, D. Mann, E.E. Levin, M. Shatruk, and R. Seshadri, Appl. Phys. Lett. **116**, 212403 (2020).

[57] L.H. Lewis, R. Barua, and B. Lejeune, J. Alloys Compd. **650**, 482 (2015).

[58] S.P. Bennett, S. Kota, H. ElBidweihy, J.F. Parker, L.A. Hanner, P. Finkel, and M.W. Barsoum, Scr. Mater. **188**, 244 (2020).

[59] S. Kota, M.T. Agne, K. Imasato, T. Aly El-Melegy, J. Wang, C. Opagiste, Y. Chen, M. Radovic, G.J. Snyder, and M.W. Barsoum, J. Eur. Ceram. Soc. **42**, 3183 (2022).

[60] I.A. Radulov, D.Y. Karpenkov, K.P. Skokov, A.Y. Karpenkov, T. Braun, V. Brabänder, T. Gottschall, M. Pabst, B. Stoll, and O. Gutfleisch, Acta Mater. **127**, 389 (2017).

[61] H. Zhang, J. Liu, M. Zhang, Y. Shao, Y. Li, and A. Yan, Scr. Mater. **120**, 58 (2016).

[62] T. Faske and W. Donner, J. Appl. Crystallogr. **51**, 761 (2018).

[63] J. Rodríguez-Carvajal, Phys. B Condens. Matter **192**, 55 (1993).

[64] K. Momma and F. Izumi, J. Appl. Crystallogr. **44**, 1272 (2011).

[65] C.A. Schneider, W.S. Rasband, and K.W. Eliceiri, Nat. Methods **9**, 671 (2012).

[66] M. Ghahremani, H.M. Seyoum, H. ElBidweihy, D.E. Torre, and L.H. Bennett, AIP Adv. **2**, 32149 (2012).

[67] V. Franco, J.S. Blázquez, B. Ingale, and A. Conde, Annu. Rev. Mater. Res. **42**, 305 (2012).

[68] L. Verger, S. Kota, H. Roussel, T. Ouisse, and M.W. Barsoum, J. Appl. Phys. **124**, 205108 (2018).

[69] C. Gianoglio and C. Badini, J. Mater. Sci. **21**, 4331 (1986).

[70] C. Kapfenberger, B. Albert, R. Pöttgen, and H. Huppertz, Zeitschrift Für Krist. **221**, 477 (2006).

[71] J. Liu, S. Li, B. Yao, J. Zhang, X. Lu, and Y. Zhou, Ceram. Int. **44**, 16035 (2018).

[72] C.L. Chien and K.M. Unruh, Phys. Rev. B **24**, 1556 (1981).

[73] O. V. Zhdanova, M.B. Lyakhova, and Y.G. Pastushenkov, Met. Sci. Heat Treat. **55**, 68 (2013).

[74] P. Chai, S.A. Stoian, X. Tan, P.A. Dube, and M. Shatruk, J. Solid State Chem. **224**, 52 (2015).

[75] L. Ke, B.N. Harmon, and M.J. Kramer, Phys. Rev. B **95**, 104427 (2017).

[76] V. Franco and A. Conde, Int. J. Refrig. **33**, 465 (2010).

[77] B.K. Banerjee, Phys. Lett. **12**, 16 (1964).

[78] B.T. Lejeune, R. Barua, Y. Mudryk, M.J. Kramer, R.W. McCallum, V. Pecharsky, and L.H. Lewis, J. Alloys Compd. **886**, 161150 (2021).

[79] Y. Bai, X. Qi, X. He, G. Song, Y. Zheng, B. Hao, H. Yin, J. Gao, and A. Ian Duff, J. Am. Ceram. Soc. **103**, 5837 (2020).

[80] D. Potashnikov, E.N. Caspi, A. Pesach, S. Kota, M. Sokol, L.A. Hanner, M.W. Barsoum, H.A. Evans, A. Eyal, A. Keren, and O. Rivin, Phys. Rev. Mater. **4**, 84404 (2020).

[81] Institut für Seltene Erden und Metalle AG, www.ise-metal-quotes.com, Date accessed: 2022-10-27.

[82] SMM Information & Technology Co Ltd., www.metal.com, Date accessed: 2022-10-27.

[83] C.J. Berlet and K. Samnani, www.mineralprices.com, Date accessed: 2022-10-27.

[84] C. Mermer and H. Şengül, J. Clean. Prod. **270**, 122192 (2020).

[85] C. Shen, Q. Gao, N.M. Fortunato, H.K. Singh, I. Opahle, O. Gutfleisch, and H. Zhang, J. Mater. Chem. A **9**, 8805 (2021).

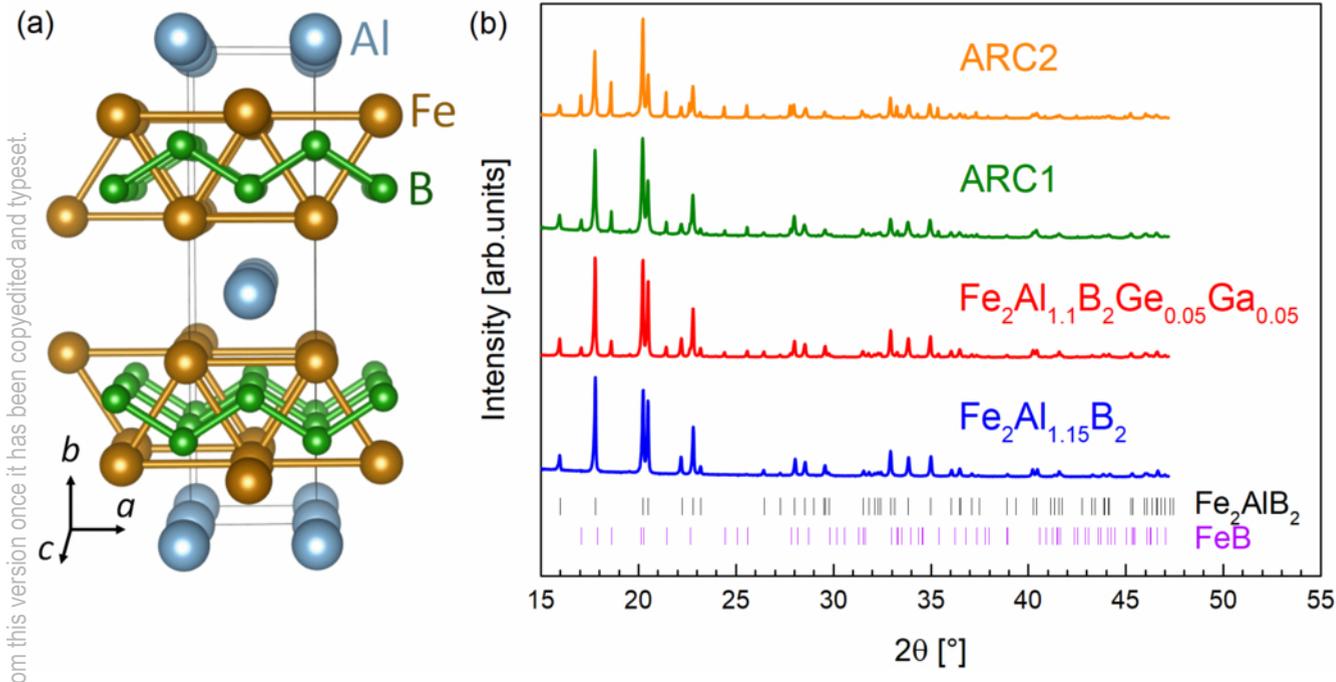
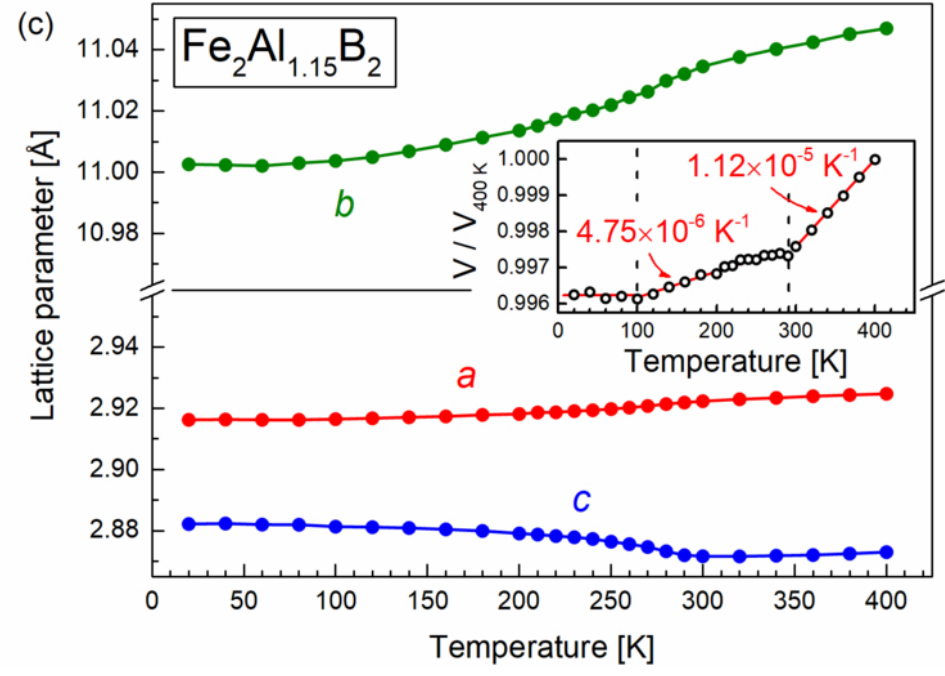

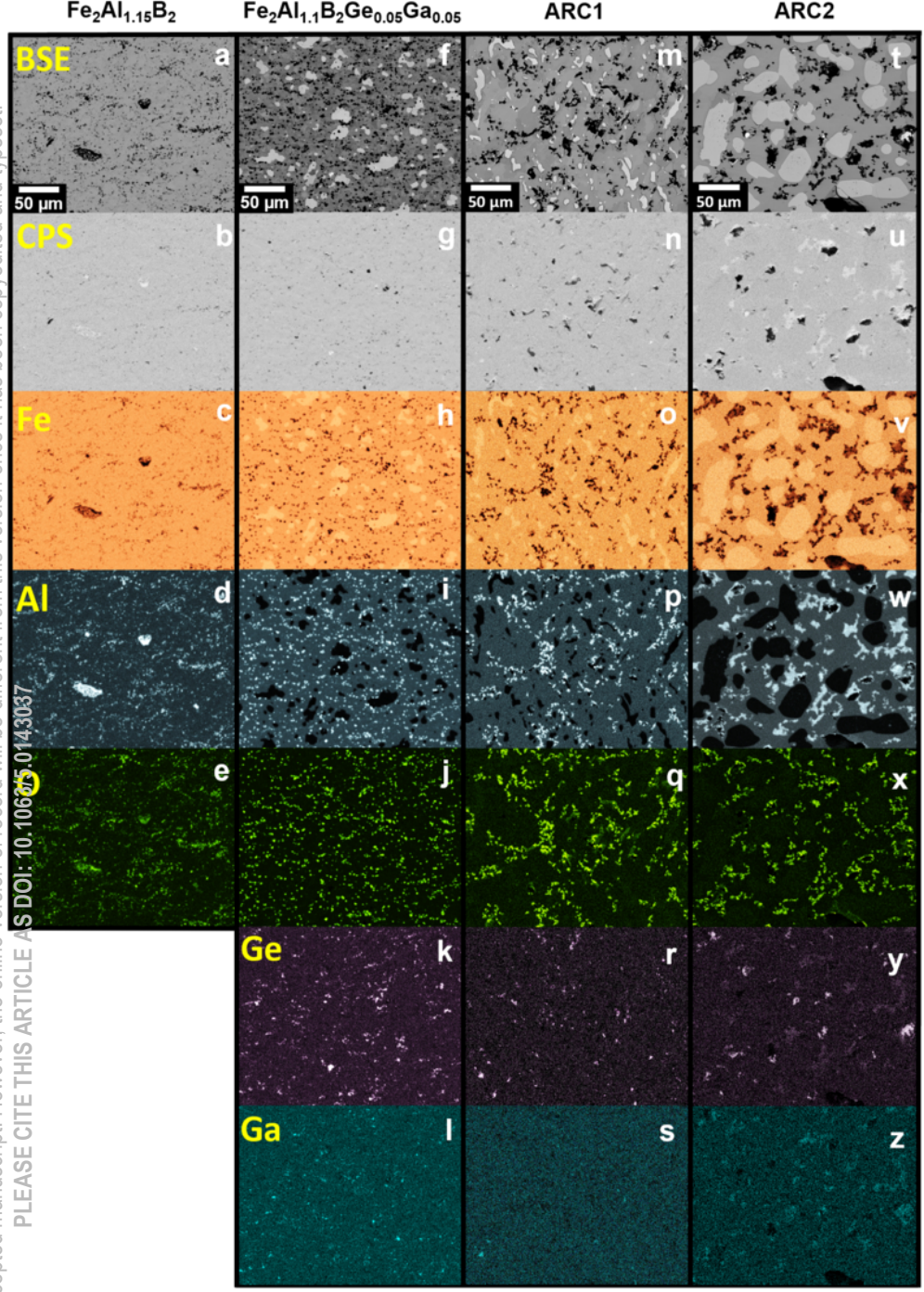

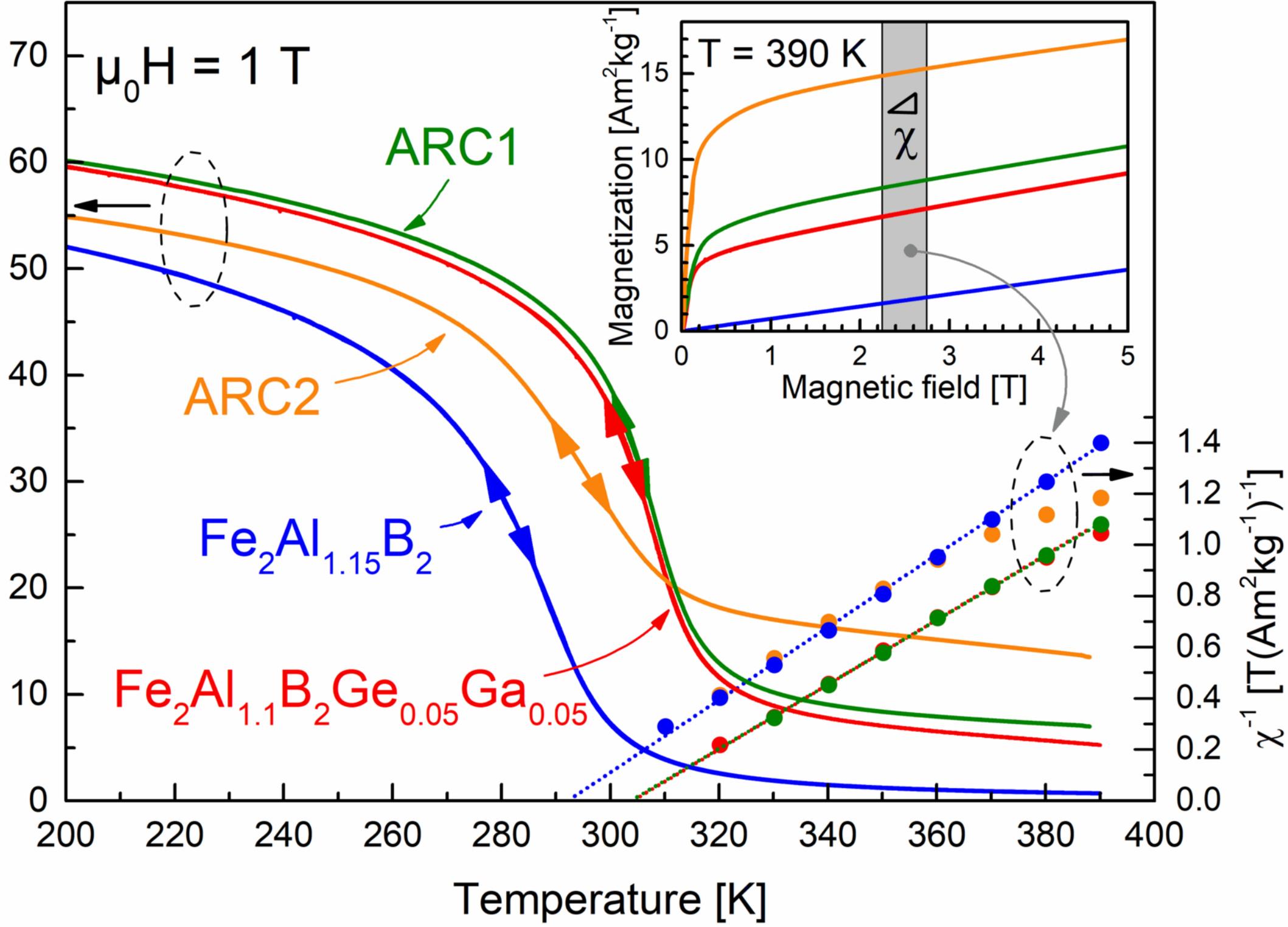

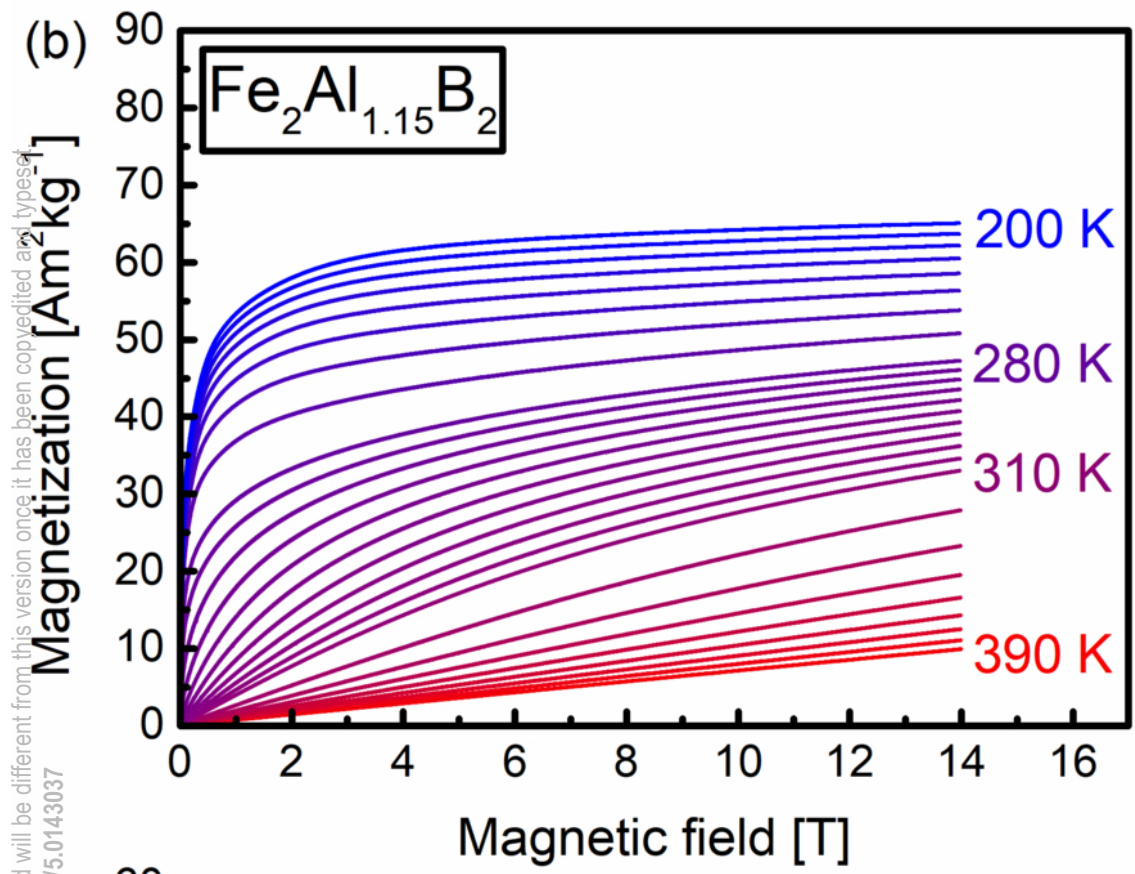
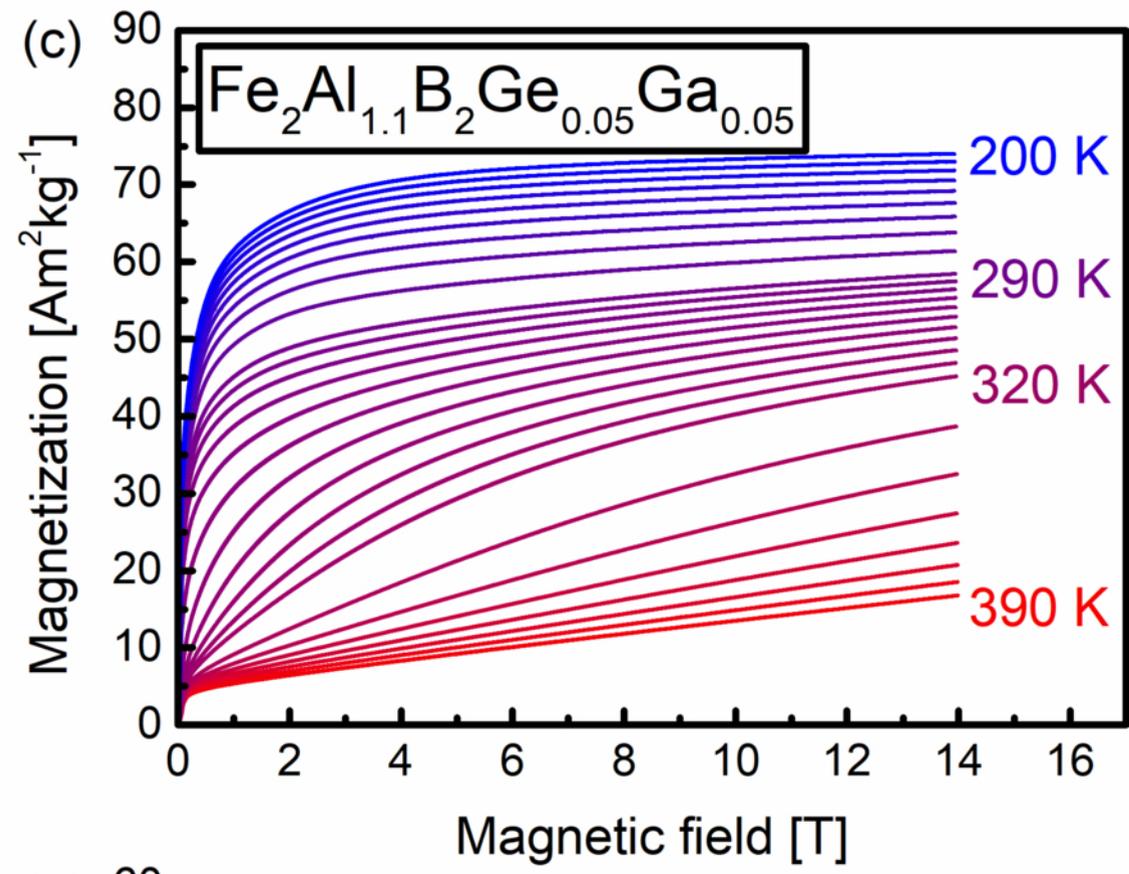
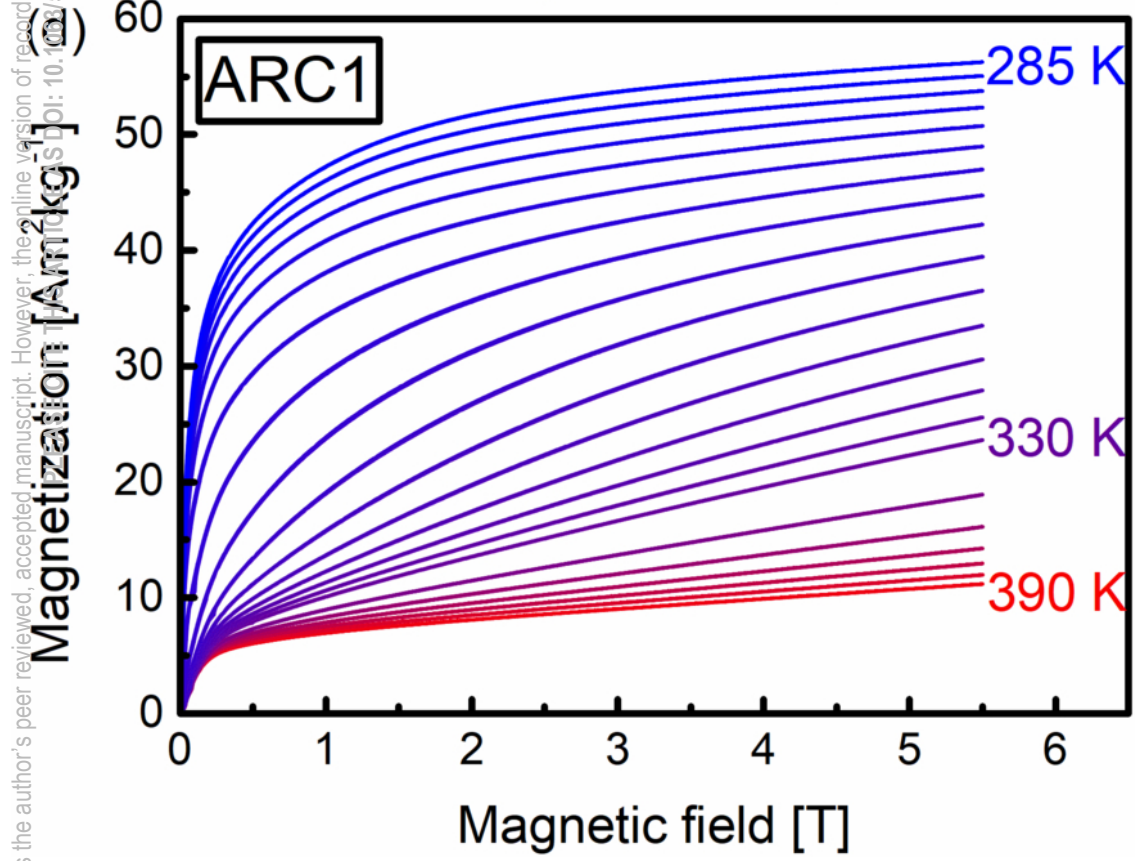
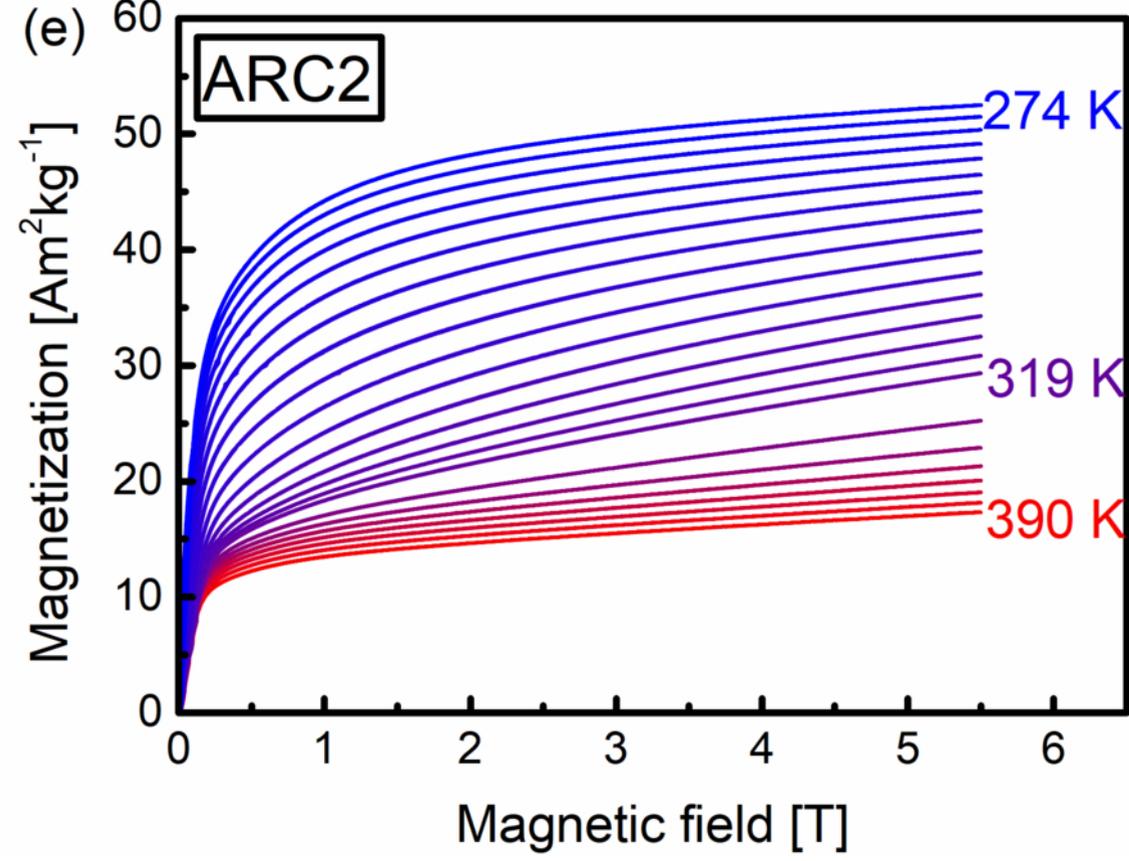

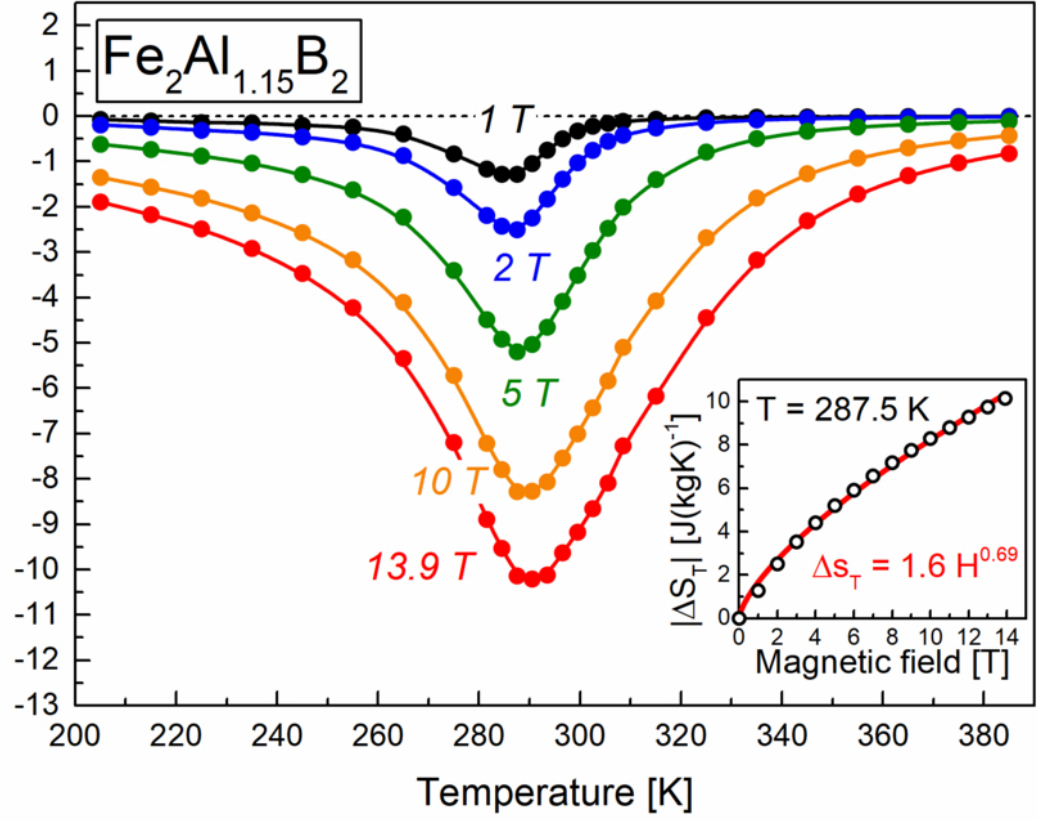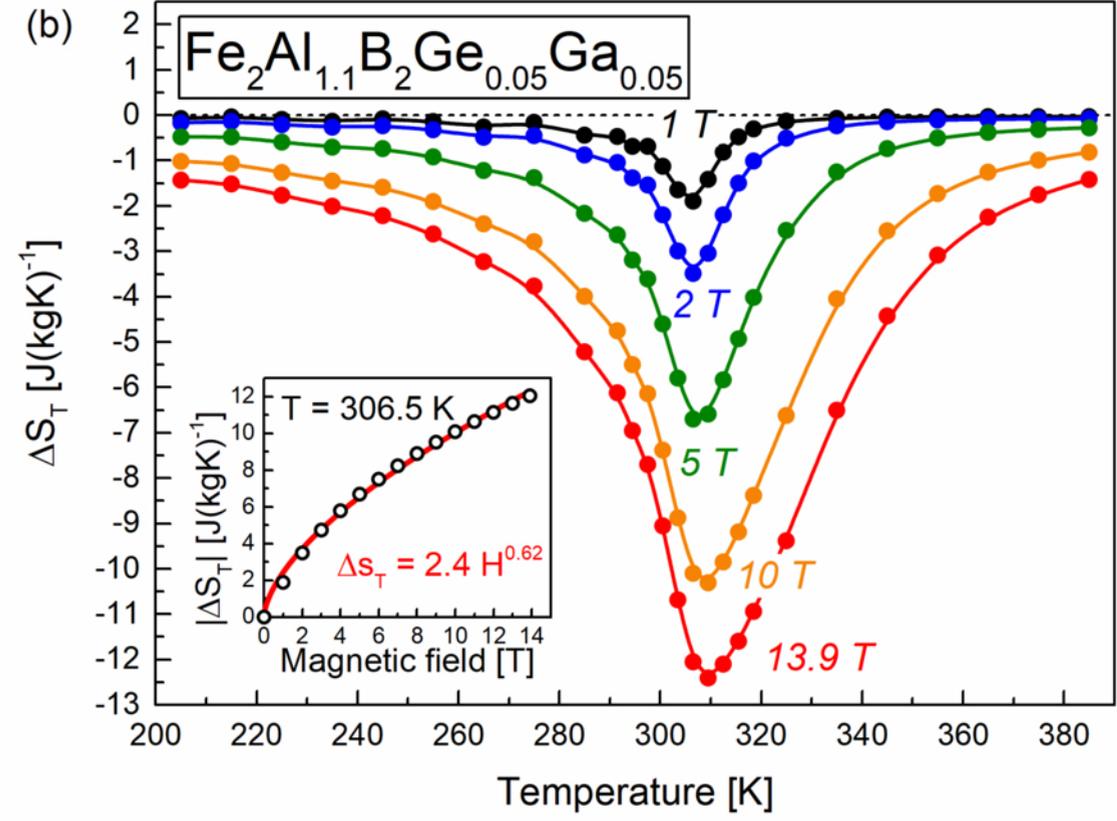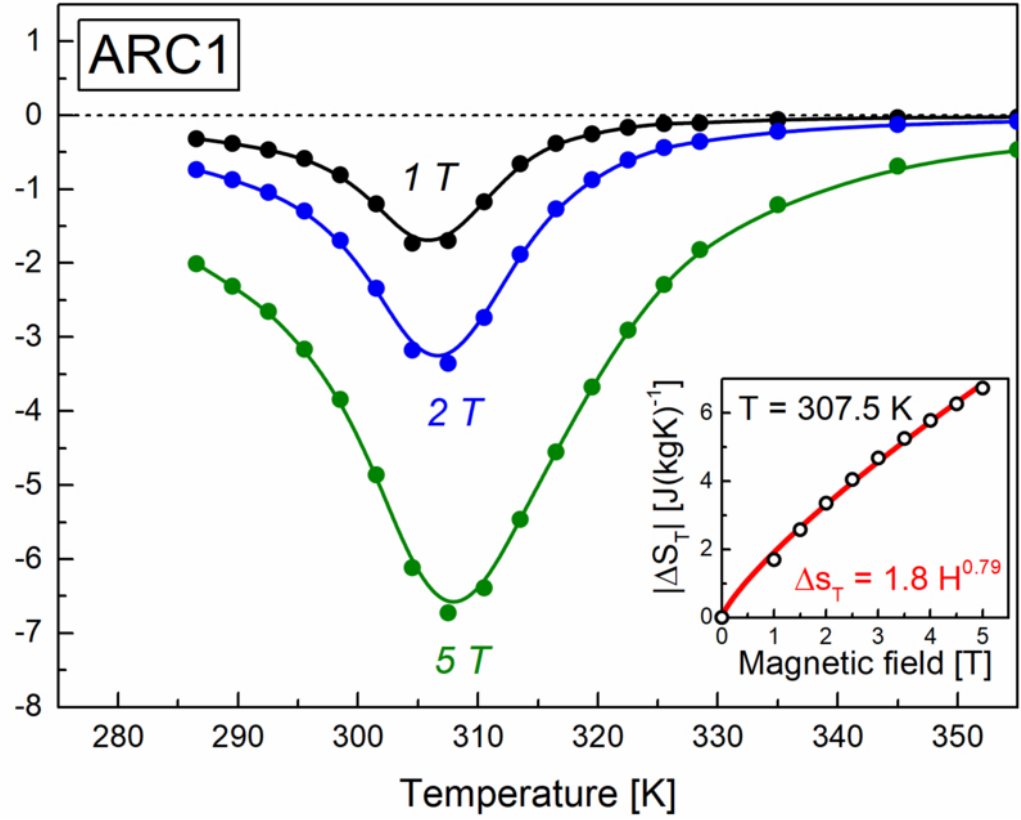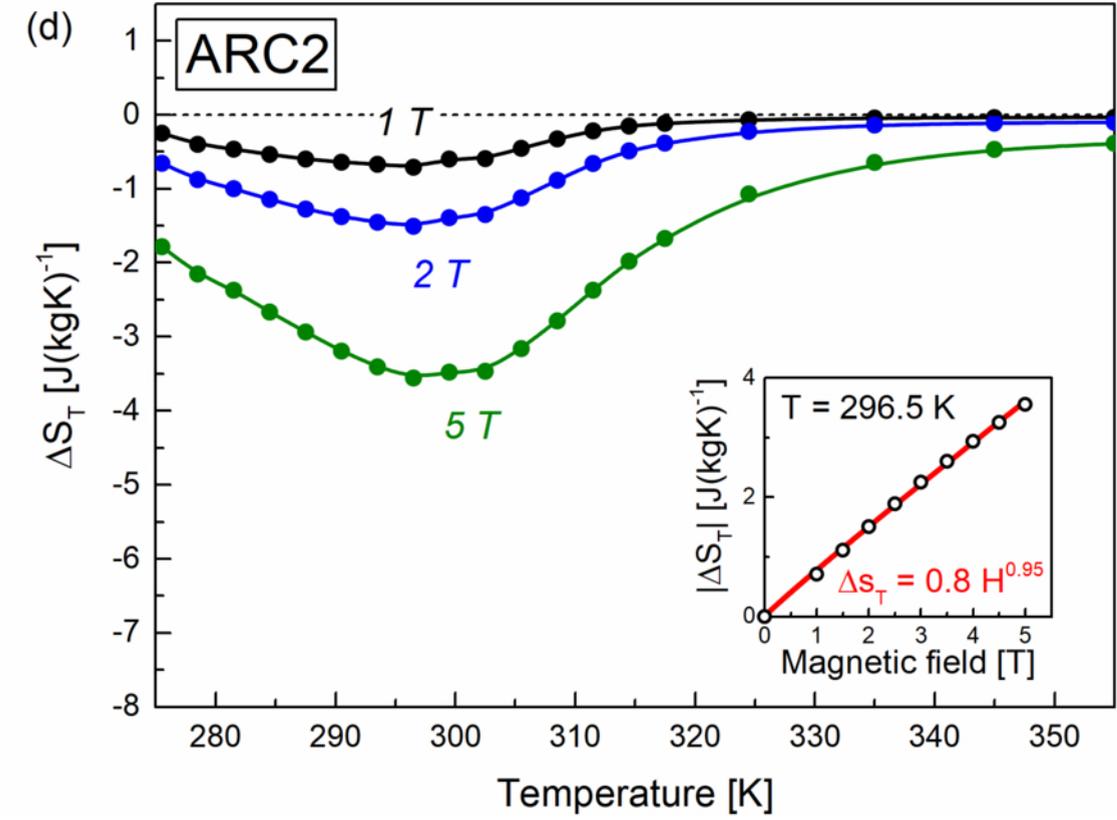

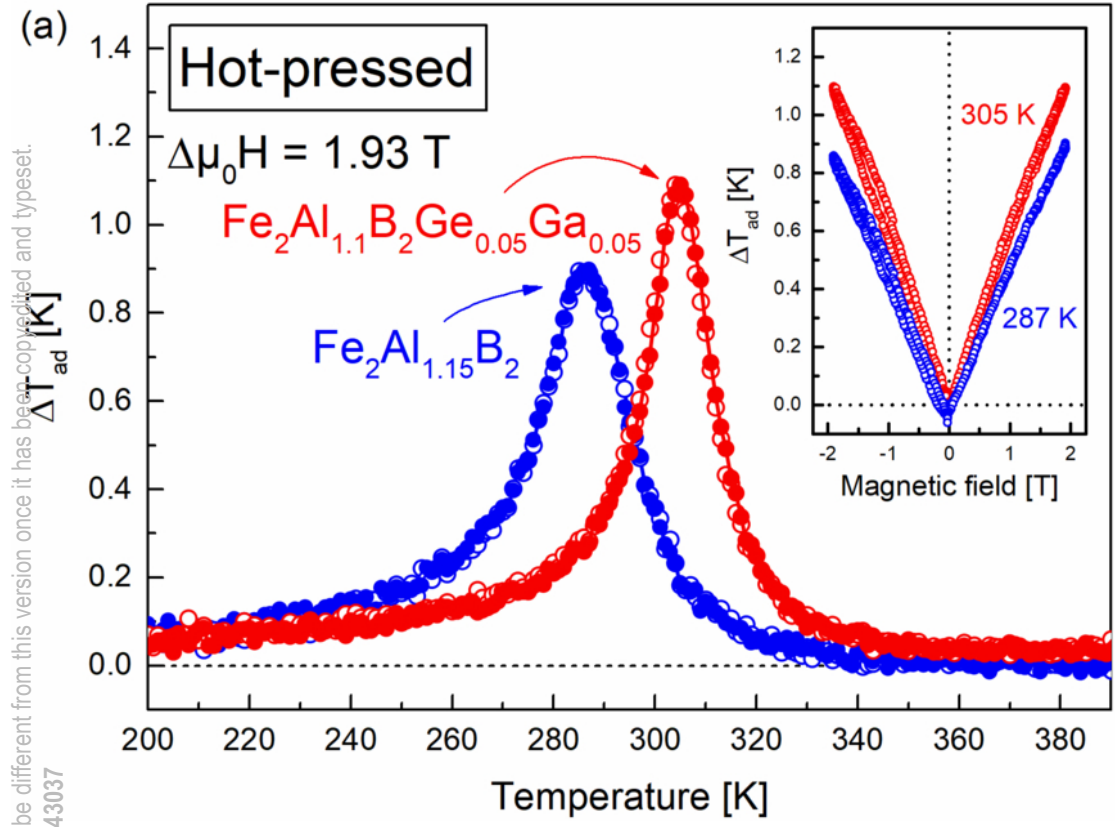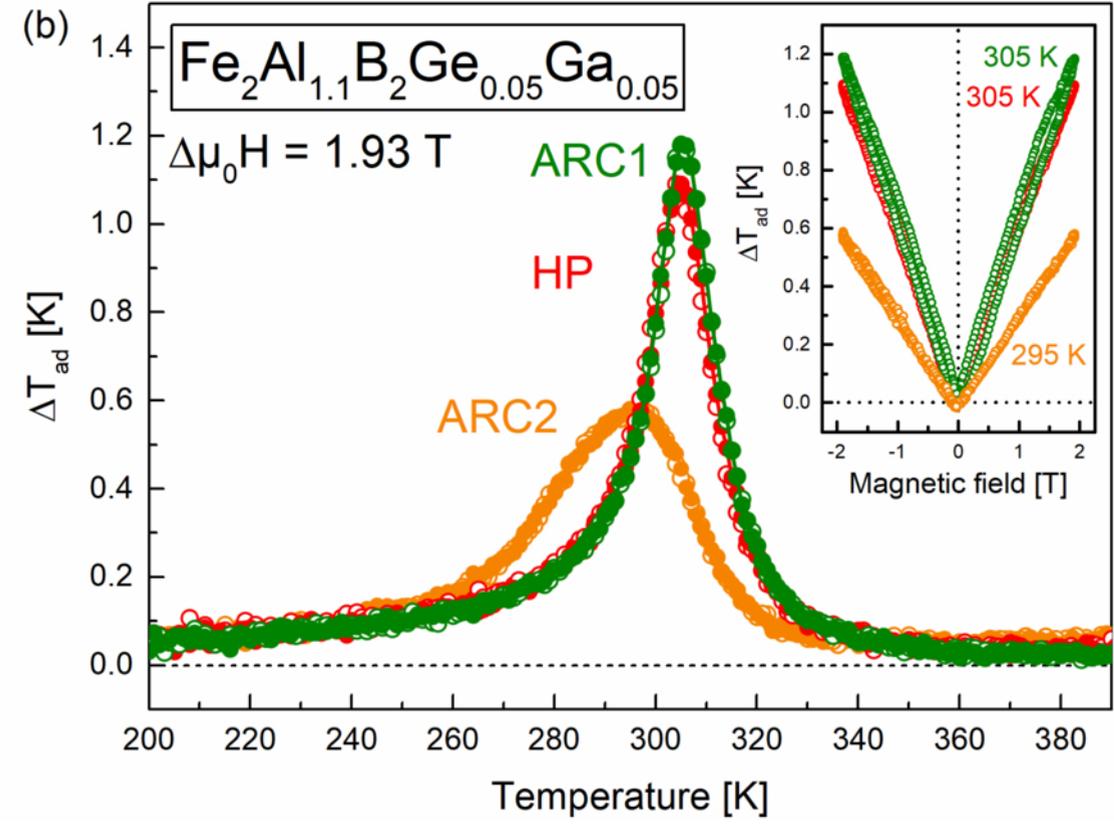

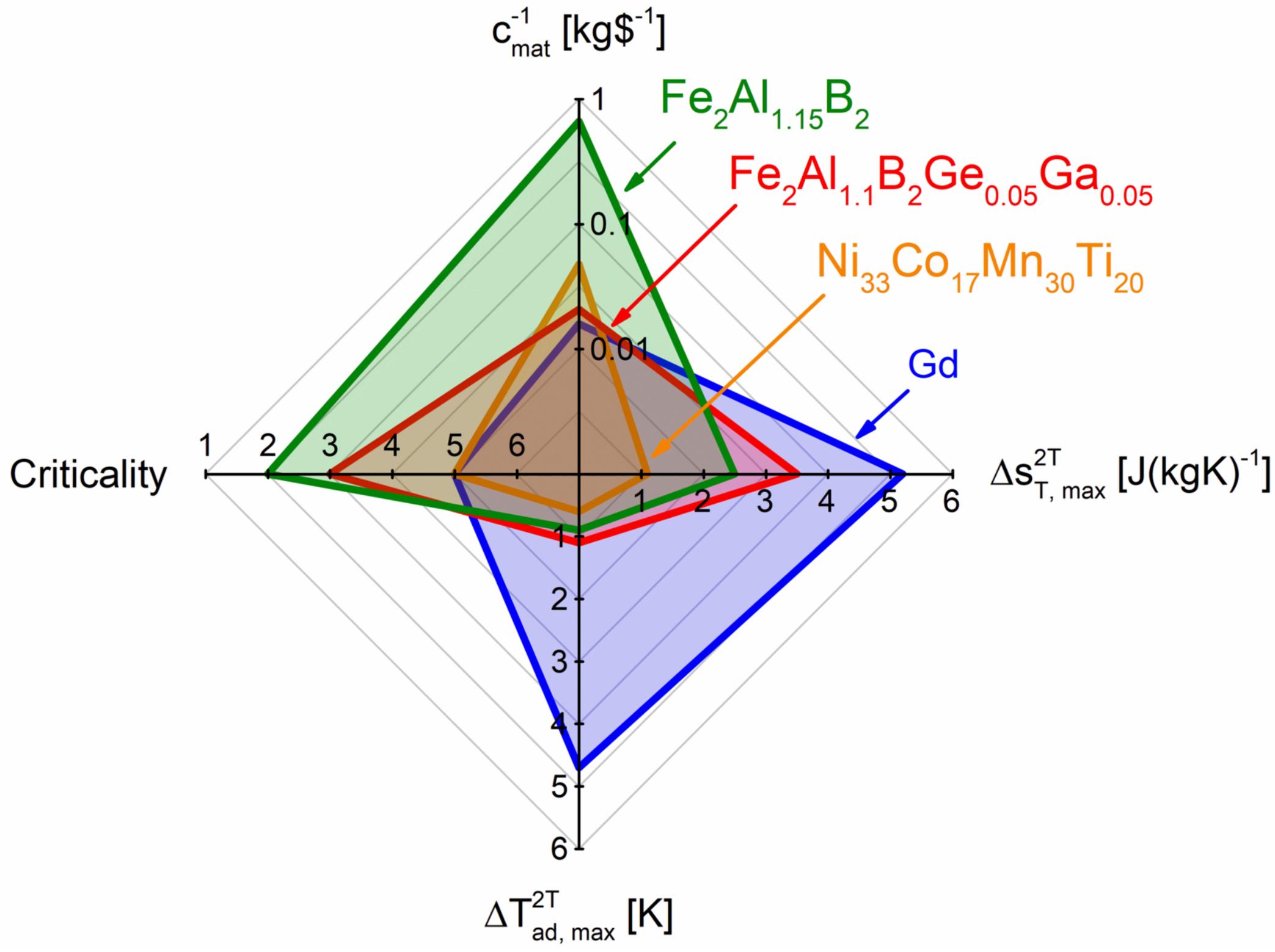